\documentstyle[12pt,a4,epsf]{article}
\newcommand{\be}{\begin{equation}}
\newcommand{\ee}{\end{equation}}
\newcommand{\ba}{\begin{eqnarray}}
\newcommand{\ea}{\end{eqnarray}}
\newcommand{\dis}{\displaystyle}
\renewcommand{\mathrm}[1]{\mbox{ #1}}
\newcommand{\tr}{\mathrm{tr}}
\def\theequation{\arabic{section}.\arabic{equation}}

\begin{document}
\begin{titlepage}
\begin{flushright}
{BERN-98/02}\\
{LU TP 98--2}\\
{UG-FT-78/97}\\
{hep-ph/9801326}
\end{flushright}
\vspace{2cm}
\begin{center}
{\Large\bf
Obtaining $K\to \pi\pi$ from Off-Shell $K\to\pi$ Amplitudes.}
\vfill
{\bf Johan Bijnens$^a$, Elisabetta Pallante$^b$, and Joaquim Prades$^{c}$}
      \\[0.5cm]
${}^a$ Department of Theoretical Physics 2, University of Lund,\\
S\"olvegatan 14A, S 22362 Lund, Sweden\\[0.5cm]
$^b$ Institut f\"ur Theoretische Physik, Universit\"at Bern,\\
Sidlerstrasse 5, CH-3012 Bern, Switzerland \\[0.5cm]
$^c$ Departamento de F\'{\i}sica Te\'orica y del Cosmos, Universidad
de Granada, Campus de Fuente Nueva, E-18002 Granada, Spain
\end{center}
\vfill
\begin{abstract}
We properly define off-shell $K\to\pi$ transition amplitudes and use them to 
extract information for on-shell $K\to\pi\pi$ amplitudes within Chiral
Perturbation Theory. At order
$p^2$ in the chiral expansion all three parameters of weak interaction 
can be determined. At order $p^4$ we are able to fix eleven additional 
constants out of thirteen contributing to off-shell  $K\to\pi$ transitions,
which leaves four undetermined constants in the on-shell $K\to\pi\pi$ 
amplitudes. All ${\cal O}(p^4)$ contributions have been exactly derived with 
 $m_\pi^2\ne 0$.   
We finally discuss the weak mass term issue and find 
contributions to on-shell $\Delta S=\pm 1$ Kaon decays,
in particular to transitions like 
$K_L \to \gamma \gamma$, $K_L \to \mu^+ \mu^-$ and
$K_S \to \pi^0 \gamma \gamma$ at the lowest non-zero order. 
\end{abstract}
\vspace*{1cm}
PACS numbers: 11.30.Rd, 13.25.Es, 12.38.Gc, 12.39.Fe

\vfill
\noindent January 1998
\end{titlepage}

\section{Introduction}
\setcounter{equation}{0}

The explanation of the $\Delta I = 1/2$ rule in
$K\to\pi\pi$ decays remains one of the challenges in Kaon physics and in
our understanding of strong interactions.
Various non-leptonic Kaon decays are also used to put limits on
CP-violation and several other quantities of the Standard Model and
extensions of it \cite{Burasreviews}.
The short distance part of the relevant operators can
be treated using renormalization group within perturbative QCD, while 
the computation of matrix elements of the relevant operators 
between meson states is a pure non perturbative problem.

In the long term lattice QCD
should be able to perform a direct computation of weak matrix elements. 
It is however much
easier on the lattice\footnote{Computing the $K\to\pi\pi$ amplitudes 
directly is quite difficult
because of the Maiani-Testa argument \cite{MT}.},
and often also in analytical attempts to reproduce the
weak matrix elements, to calculate correlators involving fewer external legs.
As a first step, current algebra can be used to relate $K\to2\pi$
to $K\to\pi$ amplitudes where it involves off-shell $K\to\pi$ transitions.
Chiral Perturbation Theory (CHPT) \cite{CHPTreviews}
is however the more modern tool to exploit the consequences
of current algebra. At lowest order in the chiral expansion 
this problem was first worked
out in \cite{Bernard} and subsequently the non-analytic parts of the 
loop contributions to  $K\to\pi\pi$ and $K\to\pi$ were
calculated in \cite{Bloops}.
It was also discussed in \cite{Yaouanc} and in the context of Wilson Fermions
on the lattice in \cite{Maiani}. 
In this paper we extend the previous study in
two ways:\\
1) We systematically go to next-to-leading order (i.e. order $p^4$) 
in CHPT and \\
2) Instead of the vague notion of off-shell Kaon and pion fields, we use
pseudo-scalar current correlators which are well
defined quantities\footnote{We have verified that in the cases discussed 
here the use of 
other two-point correlators does not yield additional information.}.

The use of this type of correlators to extract information on 
non-leptonic Kaon matrix elements is quite common in lattice studies
(see for instance \cite{Maiani}), though in those cases
 an on-shell extrapolation is usually performed. In addition,
this extrapolation is done at lowest order $p^2$ in the chiral expansion, 
i.e. using pure current algebra relations. However, due to the large Kaon 
mass, one expects non-negligible higher order CHPT corrections to Kaon weak 
matrix elements and in general to the pseudo-scalar current 
correlators involved. 
The use of
the off-shell behaviour of this type of correlators to obtain
additional information on the relevant matrix element 
has been advocated in 
\cite{BPBK}, where it was used to unravel the quark mass dependence of $B_K$,
and in \cite{BPdashen} to disentangle the structure of the
electromagnetic mass differences.

That chiral corrections are important in non-leptonic Kaon decays is already
known since a long time \cite{Bloops,pagels} and has been fully worked 
out in CHPT
by \cite{KMW1,Kambor,KMW2}. Here we present results for the
octet and the 27-plet contributions to $K\to\pi\pi$ transitions both at order 
$p^2$ and order $p^4$ and without neglecting  $m_\pi^2/m_K^2$ suppressed 
contributions.
(We have a small disagreement here with respect to previous
literature for the 27-plet \cite{KMW2}.)
While for the physical case
neglecting $m_\pi^2/m_K^2$ terms is a reasonable approximation
it will not be the case for foreseeable lattice calculations.
We keep $\overline m \equiv m_u=m_d\ne 0$ throughout the derivation.
Another issue to be clarified is the weak mass term contribution.
It is well known that the weak mass term, which gives rise to the tadpole
contributions in the original formulation of the weak effective Lagrangian 
\cite{KMW1},
 does not enter the $K\to\pi\pi$ on-shell matrix elements at
order $p^2$ \cite{Sonoda,Crewther}. It does however contribute in a well
defined way to off-shell quantities at order $p^2$ and higher. 
For this reason we shall discuss the precise role of the weak mass term 
up to order  $p^4$  in Kaon transition amplitudes in Sect. \ref{weakmass}.

It turns out that, while at order $p^2$ all the weak parameters can be
determined from our two-point correlators\footnote{These are 
called three-point correlators
in lattice QCD because of the extra weak vertex.},
this is no longer true at order $p^4$. There are in total nineteen parameters
entering the weak effective Lagrangian up to that order as discussed 
in Sect. \ref{orderp4}. Of these, we can obtain fourteen using our procedure.
Five of them can be obtained in more than one place, thus providing
as many  CHPT relations.

Several relations are also implied between different two-point
correlators so that the same quantities can also be used to check
how well calculations within CHPT obey the chiral symmetry predictions
to order $p^4$.
In particular, we can obtain several coefficients of
terms which involve quark masses at higher order. In the
purely strong sector these are the most difficult ones to predict from 
models and/or dispersive constraints. 
Determining some of them through our procedure will provide a good check 
on models used in this context (see  e.g. \cite{Antonelli}).

We study in CHPT the pseudo-scalar current correlators
\be
\label{def2point}
\Pi^{ij}(q^2) \equiv i\int \mbox{d}^4 x \, 
e^{i q\cdot x}\langle0|
T\left(P^{i\dagger}(0) P^j(x) \, 
e^{i\Gamma_{\Delta S=a}} \right)|0\rangle
\ee
in the presence of strong interactions. 
Above, $a=\pm 1, \pm 2$ stands for $\vert \Delta S\vert =1,2$ transitions
and $i,j$ are light quarks combinations 
corresponding to the octet of light pseudo-scalar mesons:
\ba
P^{\pi^0}(x) &\equiv& \frac{1}{\sqrt 2} \left(
\overline u i\gamma_5 u - \overline d i \gamma_5 d \right);\quad
P^{\pi^+}(x) \equiv \overline  d i\gamma_5 u ;\quad
P^{K^0}(x) \equiv \overline s i\gamma_5 d ;\nonumber \\ 
P^{K^+}(x) &\equiv& \overline s i\gamma_5 u ; \quad
P^{\eta_8}(x)\equiv \frac{1}{\sqrt{6}}\left(
\overline u i\gamma_5u+\overline d i\gamma_5d-
2\overline s i\gamma_5s\right)\, .
\ea
The effective action of weak interactions $\Gamma_{\Delta S=a}$ describes
strangeness changing processes in one and two units. Within the
Standard Model it can be written as follows
\be
\label{weakoper}
\Gamma_{\Delta S=a}\equiv - C_{\Delta S=a} \, 
G_F\int \rm{d}^4 y \, {\cal O}_{\Delta S=a}(y)\, ,
\ee
where ${\cal O}_{\Delta S=a}$ is a sum over the effective
operators arising after integrating out the heavy bosons,
i.e. $W$, $Z$, and the Higgs boson, and heavy fermions, 
i.e. top, bottom, and charm quark (see e.g. 
\cite{Burasreviews,Gilman-Wise,Martinelli} for the actual derivation).
The constant $C_{\Delta S=a}$ collects Clebsch-Gordan
factors and $G_F$ is the Fermi constant. 

In (\ref{def2point}), the first term in the expansion of 
$\exp [ {i\Gamma_{\Delta S=a}} ]$ 
describes strangeness zero changing transitions, 
the second term describes strangeness one and two
changing processes, while the third term includes
$(\Delta S=\pm 1)^2$ transitions.
The $\vert\Delta S\vert =2$ case relevant to the $B_K$ factor
which parameterizes the $\overline{K^0}-K^0$ mixing
was already studied in \cite{BPBK}; in Section 3,
we will just repeat the relevant expressions for completeness.
The plan of the paper is as follows. In Section 2 we construct the weak 
effective Lagrangian up to order $p^4$ and relevant to our analysis. The 
$1/N_c$ counting of the weak constants is also done in subsection 2.2.
In Section 3 the fully renormalized two-point current correlators up to 
order $p^4$ are derived for $\vert\Delta S\vert =0,1,2$ cases. 
The non-analytic 
contributions to one loop are collected in Appendix \ref{applogs1}. 
In Section 4 the $K\to \pi\pi$ on-shell amplitudes are derived in CHPT 
up to order $p^4$. In Appendix \ref{applogs2} are the non-analytic 
contributions to one loop. Section \ref{kptokpp} is devoted to the connection
between off-shell $K\to \pi$ transition amplitudes and on-shell
$K\to 2\pi$    
amplitudes.
Resonance saturation also for the 27-plet sector is used here and derived
in Appendix \ref{appres}. 
Finally, in Section \ref{weakmass} we clarify the role of the 
weak mass term in Kaon decays up to order $p^4$ and in Section 
\ref{conclusions} we state our conclusions. 

\section{CHPT Lagrangian and $1/N_c$-Discussion}
\setcounter{equation}{0}

At lowest order in CHPT (i.e. ${\cal O}(p^2)$) the strangeness changing
interactions up to two units 
amongst the pseudo-Goldstone bosons and external
scalar, pseudo-scalar, vector and axial-vector sources 
(neglecting virtual photon interactions) are
described by the following effective Lagrangian:
\ba
{\cal L}^{(2)}_{\mbox{\tiny{eff}}} &=& {\cal L}^{(2)}_{\Delta S=0} 
+ {\cal L}^{(2)}_{\Delta S=1}   
+ {\cal L}^{(2)}_{\Delta S=2}   
\, .
\ea
The first term is the strong interaction Lagrangian
\ba
\label{lowestorder}
{\cal L}^{(2)}_{\Delta S=0} &=& \frac{F_0^2}{4}\,  \left[
\tr \left(u^\mu u_\mu\right)+ \tr \left( \chi_+ \right) \right]\, ,
\ea
where $\tr (A)$ is the flavour trace of $A$,  
\be
u_\mu\equiv i u^\dagger \left(D_\mu U\right) u^\dagger = u_\mu^\dagger
\ee
and $U\equiv u^2 = \exp(i \sqrt 2 \Phi/F_0)$ is the exponential representation
 incorporating the  SU(3) matrix of the octet of light pseudo-scalar mesons  
\be 
\label{Uoctet}
\Phi(x)=\frac{\vec{\lambda}\cdot \vec{\phi} }{\sqrt 2} =
\left( \begin{array}{ccc}
\frac{\dis \pi^0}{\dis \sqrt 2} + \frac{\dis \eta_8}
{\dis \sqrt 6} &  \pi^+ & K^+ \\
\pi^- & -\frac{\dis \pi^0}{\dis \sqrt 2}+\frac{\dis \eta_8}
{\dis \sqrt 6} & K^0 \\
K^- & \overline K^0 & -\frac{\dis 2 \eta_8}{\dis \sqrt 6} \end{array}
\right) \, .
\ee
 $D_\mu U$  denotes the covariant derivative on the $U$ field
\be
\label{defDU}
D_\mu U \equiv \partial_\mu
U -i (v_\mu+a_\mu)U+iU(v_\mu-a_\mu) \, ,
\ee
where $v_\mu(x)$ and $a_\mu(x)$ are external SU(3)
vector and axial-vector matrices (no singlet component will be included 
in the present analysis). The matrix $\chi_+$ in 
Eq. (\ref{lowestorder}) and its pseudoscalar 
counterpart $\chi_-$ are defined as follows 
\ba
\chi_{\pm}&\equiv&u^\dagger \chi u^\dagger \pm u \chi^\dagger u\, ,
\ea
where $\chi\equiv 2 B_0 ({\cal M}+s(x)+ip(x))$, 
$s(x)$ and $p(x)$ are external scalar and pseudo-scalar SU(3) 
sources and ${\cal M}\equiv\mbox{ diag}
(m_u,m_d,m_s)$ is the light quarks mass matrix.
The constant $B_0$ is related to
the vacuum expectation value of the scalar quark density
\ba
\langle 0 | \overline q q | 0 \rangle\left|_{q=u,d,s}
\right. \equiv -F_0^2 B_0 \left(1+{\cal O(M)}\right) \, .  
\ea
With this normalization, $F_0$ is the chiral limit value 
of the pion decay constant $F_\pi \simeq
92.4$ MeV. In the absence of the U(1)$_A$ anomaly (i.e. in the large $N_c$
limit) \cite{tHooft}, the U(3) singlet field $\eta_1$ becomes the
ninth Goldstone boson which is incorporated in the $\Phi(x)$ field as
\be
\label{eta1}
\Phi(x)= \frac{\vec{\lambda}\cdot\vec{\phi} }{\sqrt 2}+ \frac{\eta_1}{\sqrt 3}
{\mbox {\large \bf 1}} \, .
\ee
In this work we limit ourselves to the octet symmetry case, meaning that
we assume
the singlet degree of freedom $\eta_1$ as heavy and integrated out.
This is enough for our purpose of showing how to relate 
off-shell $K \to \pi$ and $K \to \eta_8$ transitions 
to on-shell $K \to \pi \pi$ amplitudes. 
In octet symmetry, i.e. $\det (u)= 1$ and $\tr (u_\mu)=0$,
the $\Delta S=\pm 1$ contribution  to the l.h.s. of (\ref{weakoper}) 
is given by
\ba
\label{deltas1}
{\cal L}_{\Delta S=1}^{(2)} \;=\;& C \, F_0^4& \, 
\Bigg[ G_8 \tr \left( \Delta_{32} u_\mu u^\mu \right) +
G_8' \tr \left(\Delta_{32} \chi_+ \right) \nonumber \\&& +
G_{27} t^{ij,kl} \, \tr \left(\Delta_{ij} u_\mu \right)
\tr \left( \Delta_{kl} u^\mu \right) \Bigg]
+ \mbox{ h.c.} \, ,
\ea
where $i,j,k,l=1,2,3$ correspond to the light flavour indices $u, d, s$ and the
tensor $t^{ij,kl}$ has
\ba
\label{deft}
t^{21,13} &=& 
t^{13,21} = \frac{1}{3} \, ; \, 
t^{22,23}=t^{23,22}=-\frac{1}{6} \, ; \nonumber \\
t^{23,33}&=&t^{33,23}=-\frac{1}{6} \, ; \, 
t^{23,11} =t^{11,23}=\frac{1}{3} \, 
\ea
and zero otherwise.
The matrix $\Delta_{ij}$ is defined as
\ba
\Delta_{ij} &\equiv& u \lambda_{ij} u^\dagger \nonumber \\
\left(\lambda_{ij}\right)_{ab} &\equiv& \delta_{ia} \, \delta_{jb} \, , 
\ea
with $\delta_{ia}$ the Kronecker delta acting on the SU(3)
light flavour space.
The constant $C=C_{\Delta S=1}$ in (\ref{weakoper})
includes normalization factors and the
Cabibbo-Kobayashi-Maskawa matrix elements
\be
C= -\frac{3}{5} \, \frac{G_F}{\sqrt 2} V_{ud} \, V_{us}^* \, .
\ee
The couplings $G_8$ and $G_8'$ modulate octet operators under
SU(3)$_L$ $\times$ SU(3)$_R$, while
$G_{27}$ modulates a 27-plet operator. For on-shell 
$K \to \pi \pi$ transitions
at order $p^2$ one can set $G_8'=0$ \cite{Bernard,Sonoda,Crewther}.
 At order $p^4$ this question was studied
in \cite{KMW1,Leurer}; the result is that one can always use a basis
for the order $p^4$ counterterms where the effects of the weak mass term 
($G_8'$) on the on-shell $K \to \pi \pi$ amplitudes are fully reabsorbed
at this order. 
 Clearly, the use of the shifted basis implies a redefinition 
of the order $p^4$ couplings in order to absorb the weak mass term 
contributions. This was not done in \cite{Antonelli}.
Since $G_8'$
does always appear in off-shell $K\to \pi$ transitions, 
we keep the unshifted basis in 
our analysis, where  $K \to \pi \pi$ amplitudes explicitly contain 
order $p^4$ contributions proportional to $G_8'$. 

The $\vert\Delta S\vert =2$ term in Eq. (\ref{weakoper})
can be written as follows   
\ba
\label{deltas2}
{\cal L}_{\Delta S=2}^{(2)} &=& C_{\Delta S=2} \, F_0^4 \, 
G_{27}  \, \tr \left(\Delta_{32} u_\mu \right)
\tr \left( \Delta_{32} u^\mu \right) + \mbox{ h.c.} \, , 
\ea
with 
\ba
C_{\Delta S=2} = - \frac{G_F}{4} {\cal F }(m_t^2, m_c^2, M_W^2, 
{\bf V}_{\rm CKM}) \,
\ea
and ${\cal F }(m_t^2, m_c^2, M_W^2, {\bf V}_{\rm CKM})$ being 
a known function of the heavy fermions and bosons masses, and the
 Cabibbo-Kobayashi-Maskawa matrix elements \cite{Burasreviews}.

The weak couplings $G_8$, $G_8'$ and $G_{27}$ in Eqs. (\ref{deltas1}) and 
(\ref{deltas2}) are dimensionless and they are related to those
used in \cite{KMW1} as follows
\be
C \, F_0^4 \, G_8=c_2 \; \quad C \, F_0^4 \, G_{27} = 3 c_3 
\, ;
\quad C \, F_0^4 \, G_8' = c_5 \, .
\ee 

\subsection{The order $p^4$}
\label{orderp4}

At next-to-leading order in the chiral expansion (i.e. at order $p^4$) 
the complete
list of counterterms in the strong interaction sector and 
in the octet symmetry case has been given by 
Gasser and Leutwyler in \cite{GL1}. 
In the SU(3) flavour case and for on-shell Green's functions there appear
ten counterterms denoted with $L_i,\, i=1,\ldots,10$, while in the off-shell 
case there are two extra contact terms 
$H_1$ and $H_2$ involving external sources only.

The complete basis of counterterms in the weak interaction sector 
at order $p^4$ and  
describing transitions with strangeness changing in one and two units
was first derived by Kambor, Missimer and Wyler in \cite{KMW1}
and in \cite{EC}.
This basis was confirmed and
reduced to various minimal sets by Esposito-Far\`ese for the
octet and 27-plet operators \cite{EF}, and by Ecker, Kambor 
and Wyler for the octet operators \cite{EKW}.

For the analysis of the $\Delta S=\pm 1$ transitions 
we use a minimal set of operators which
differs from the one in \cite{EF} in one octet
operator, but has the advantage of producing shorter expressions.
Our octet subset coincides with that of Ecker et al. in \cite{EKW} up to two 
 operators.
We also give below the translation from the counterterms we are using to those
in \cite{EKW}.

In the octet symmetry case, a minimal set of counterterms  
contributing to 
the $K \to \pi$ and $K \to \pi \pi$ transitions at order $p^4$ in CHPT 
is given by:
\ba
\label{lagDS1}
{\cal L}^{(4)}_{\Delta S=1} &=& C \, F_0^2 \, G_8 \left[
E_1 {\cal O}^8_1 + E_2 {\cal O}^8_2 + E_3 {\cal O}^8_3 + 
E_4 {\cal O}^8_4 + E_5 {\cal O}^8_5 \right. \nonumber \\
&+& \left. E_{10} {\cal O}^8_{10} + 
E_{11} {\cal O}^8_{11} + E_{12} {\cal O}^8_{12} +
E_{13} {\cal O}^8_{13} + E_{15} {\cal O}^8_{15} \right]
\nonumber \\
&+& C \, F_0^2 \, G_{27} \left[
D_1 {\cal O}^{27}_1 + D_2 {\cal O}^{27}_2 +
D_{4} {\cal O}^{27}_4 + D_{5} {\cal O}^{27}_5 
\right. \nonumber \\ &+& \left. 
D_{6} {\cal O}^{27}_6 + D_{7} {\cal O}^{27}_7 \right] 
+ \mbox{h.c.} \, .  
\ea
The octet operators above are 
\ba
\label{8oper}
{\cal O}^8_1 
&=& \tr \left(\Delta_{32} \chi_+ \chi_+ \right) \, ; 
\nonumber \\
{\cal O}^8_2 
&=& \tr \left( \Delta_{32} \chi_+ \right) 
\tr \left( \chi_+  \right) \, ;  \nonumber \\
{\cal O}^8_3
&=& \tr \left( \Delta_{32} \chi_- \chi_- \right) \, ; 
\nonumber \\
{\cal O}^8_4
&=& \tr \left( \Delta_{32} \chi_- \right)\tr \left( \chi_- \right) \, ;
\nonumber \\
{\cal O}^8_5
&=& \tr \left( \Delta_{32} \left[\chi_+,\chi_-\right] \right) \, ; 
\nonumber \\ 
{\cal O}^8_{10} 
&=& \tr \left( \Delta_{32} \left\{ \chi_+, u^\mu u_\mu \right\}
 \right) \, ; \nonumber \\
{\cal O}^8_{11} 
&=&  \tr \left( \Delta_{32} u_\mu \chi_+ u^\mu \right) \, ; 
\nonumber \\
{\cal O}^8_{12} 
&=& \tr \left( \Delta_{32} u_\mu \right) 
\tr \left( \left\{ u^\mu , \chi_+ \right\} \right) \, ; \nonumber \\
{\cal O}^8_{13} 
&=& \tr \left( \Delta_{32} \chi_+ \right) \tr \left( u^\mu u_\mu 
\right) \, ; \nonumber \\
{\cal O}^8_{15}
&=& \tr\left( \Delta_{32} \left[ \chi_-, u^\mu u_\mu \right] 
\right) \, .
\ea
The 27-plet operators  are
\ba
\label{27oper}
{\cal O}^{27}_1 
&=& t^{ij,kl} \, 
\tr \left( \Delta_{ij} \chi_+\right) 
\tr\left( \Delta_{kl} \chi_+ \right) \, ;
 \nonumber  \\
{\cal O}^{27}_2 &=& t^{ij,kl} \, 
\tr \left( \Delta_{ij} \chi_-\right) 
\tr \left( \Delta_{kl} \chi_- \right) \, ;
 \nonumber  \\
{\cal O}^{27}_4 &=& t^{ij,kl} \, 
\tr \left( \Delta_{ij} u_\mu \right) 
\tr \left( \Delta_{kl} \left\{ u^\mu, \chi_+\right\} \right) \, ;
 \nonumber  \\
{\cal O}^{27}_5 &=& t^{ij,kl} \, 
\tr \left( \Delta_{ij} u_\mu \right) 
\tr \left( \Delta_{kl} \left[ u^\mu, \chi_-\right] \right) \, ;
 \nonumber  \\
{\cal O}^{27}_6 &=& t^{ij,kl} \, 
\tr \left( \Delta_{ij} \chi_+\right) 
\tr \left( \Delta_{kl} u_\mu u^\mu \right) \, ;
 \nonumber  \\
{\cal O}^{27}_7 &=& t^{ij,kl} \, 
\tr \left( \Delta_{ij} u_\mu \right) 
\tr \left( \Delta_{kl} u^\mu \right) \tr \left( \chi_+ \right) \, ,
\ea
where the tensor $t^{ij,kl}$ is defined in (\ref{deft}).

The translation from the octet counterterms in (\ref{8oper}) to 
the ones used by Ecker et al. \cite{EKW} is as follows
\ba
N_5&=& E_{10}-E_{11} \, ; ~~\quad N_6= E_{11} + 2 E_{12} \, ; \nonumber \\
N_7&=&\frac{1}{2} E_{11} + E_{13} \, ; \quad N_8= E_{11} \, ; \nonumber \\
N_9&=&E_{15} \, ; ~~~~~~~\quad N_{10} = E_1-E_5 \, ; \nonumber \\
N_{11}&=&E_2 \, ; ~~~~~~~~\quad N_{12}= -E_3 + E_5 \, ; \nonumber \\
N_{13}&=&-E_4 \, ; ~~~~~\quad N_{36}=E_5 \, .
\ea
In octet symmetry and at the same order in CHPT, 
a minimal set of counterterms contributing to 
 the $\Delta S=2$ component of
the $K^0 -\overline{K^0}$  mixing is:
\ba
\label{lagDS2}
{\cal L}^{(4)}_{\Delta S=2} &=&  C \, F_0^2 \, G_{27} \left[
D_1 {\cal O}^{\Delta S=2}_1 + D_2 {\cal O}^{\Delta S=2}_2 +
D_{4} {\cal O}^{\Delta S=2}_4 + D_{5} {\cal O}^{\Delta S=2}_5 
\right. \nonumber \\ &+& \left.  
D_{6} {\cal O}^{\Delta S=2}_6 + D_{7} {\cal O}^{\Delta S=2}_7 \right] 
+ \mbox{h.c.}\, .
\ea
Notice that this basis is not exactly the one  used
in \cite{BPBK}; the one in (\ref{lagDS2}) is a minimal set.
The $\Delta S=2$ operators are given by
\ba
\label{S=2oper}
{\cal O}^{\Delta S=2}_1 
&=& \tr \left( \Delta_{32} \chi_+\right) 
\tr\left( \Delta_{32} \chi_+ \right) \, ;
 \nonumber  \\
{\cal O}^{\Delta S=2}_2 &=& \tr \left( \Delta_{32} \chi_-\right) 
\tr \left( \Delta_{32} \chi_- \right) \, ;
 \nonumber  \\
{\cal O}^{\Delta S=2}_4 &=& \tr \left( \Delta_{32} u_\mu \right) 
\tr \left( \Delta_{32} \left\{ u^\mu, \chi_+\right\} \right) \, ;
 \nonumber  \\
{\cal O}^{\Delta S=2}_5 &=& \tr \left( \Delta_{32} u_\mu \right) 
\tr \left( \Delta_{32} \left[ u^\mu, \chi_-\right] \right) \, ;
 \nonumber  \\
{\cal O}^{\Delta S=2}_6 &=& \tr \left( \Delta_{32} \chi_+\right) 
\tr \left( \Delta_{32} u_\mu u^\mu \right) \, ;
 \nonumber  \\
{\cal O}^{\Delta S=2}_7 &=& \tr \left( \Delta_{32} u_\mu \right) 
\tr \left( \Delta_{32} u^\mu \right) \tr \left( \chi_+ \right) \, .
\ea
Since the 27-plet operators with $\Delta S=1$ in  (\ref{lagDS1})
and the $\Delta S=2$ ones above are components
of the same irreducible tensor under  
SU(3)$_L$ $\times$ SU(3)$_R$, the $D_i$ couplings
in both Lagrangians have to be the same. 

The divergences associated with the minimal set of counterterms in 
(\ref{lagDS1}) and (\ref{lagDS2})
can be extracted from Kambor et al. \cite{KMW1} with the use of the strong
equation of motion, partial integration and the Cayley-Hamilton theorem.
We explicitly verified the results in \cite{KMW1} and differ in the 27-plet 
sector by an overall sign\footnote{EP has independently redone 
the Generating Functional calculation
of the infinities and agrees with \cite{EKW,EF} and \cite{KMW1} modulo the 
27-plet overall sign.}.
The subtraction procedure is defined in the usual manner by  
\ba
E_i \equiv E_i^r + \frac{\nu^{d-4}}{16 \pi^2} \left\{
\frac{1}{d-4}+\frac{1}{2}\left[ \gamma_E-1-\ln(4\pi)\right]\right\}
\, \left[ \varepsilon_i + \frac{G_8'}{G_8} \varepsilon'_i\right] \, 
\ea
and
\ba
D_i \equiv D_i^r + \frac{\nu^{d-4}}{16 \pi^2} \left\{
\frac{1}{d-4}+\frac{1}{2}\left[ \gamma_E-1-\ln(4\pi)\right]\right\}
\, \gamma_i\, .
\ea
In the strong sector we need the counterterms 
\ba
L_i \equiv L_i^r + \frac{\nu^{d-4}}{16 \pi^2} \left\{
\frac{1}{d-4}+\frac{1}{2}\left[ \gamma_E-1-\ln(4\pi)\right]\right\}
\, \Gamma_i \, ,
\ea
with $i=$ 4, 5, 6, 7, 8,  and the one associated with the contact term
\ba
H_j \equiv H_j^r + \frac{\nu^{d-4}}{16 \pi^2} \left\{
\frac{1}{d-4}+\frac{1}{2}\left[ \gamma_E-1-\ln(4\pi)\right]\right\}
\, {\delta_j} \, , 
\ea
with  $j=2$.
The coefficients of the divergent parts are fixed to be
\be
\Gamma_4=\frac{1}{8} , \quad  \Gamma_5= \frac{3}{8} ,  \quad 
\Gamma_6= \frac{11}{144} , \quad  \Gamma_7= 0 , \quad  
\Gamma_8= \frac{5}{48} , \quad  
\mbox{ and} \quad {\delta_2}= \frac{5}{24}\,  , 
\ee
and those of Table \ref{table1}.
\begin{table}[htb]
\begin{center}
\begin{tabular}{|c|c|c|c||c|c|c|}
\hline
$E_i$ & $\varepsilon_i$ & $\varepsilon'_i$ &$N_c$& $D_i$ & $\gamma_i$&$N_c$\\
\hline 
&&&&&&\\
1& 1/4 & 5/6&1& 1 &$-$1/6&1\\
2& $-$13/18&11/18&$-2/3 E_1+{\cal O}(1/N_c)$& 2 &0&1\\
3& 0 & 0&1&4&3&$N_c$\\
4& 0& 0 &1& 5&1&1\\
5&$-$5/12& 5/12 &1&6 &$-$3/2&1\\
10& 1& 3/4&$N_c$& 7&1&1\\
11& $-$1/2& 0 &$N_c$&-- &--&--\\
12& 1/8 & 0 &$N_c$&-- &--&--\\
13& $-$7/8& 1/2&1&-- &--&--\\
15& 3/4& $-$3/4&1&-- &--&--\\
\hline
\end{tabular}
\end{center}
\protect\caption{The divergences and the leading in $1/N_c$ behaviour
of the weak ${\cal O}(p^4)$ 
octet counterterms $E_i$ and 27-plet counterterms $D_i$.
Notice that $F_0^2$ (i.e. an $N_c$ factor) is factored out.}   
\label{divergences}
\label{table1}
\end{table}

\subsection{$1/N_c$ Counting}
\label{Nccounting}

It is also useful to know the $1/N_c$ counting
of the different weak couplings in the Lagrangians 
(\ref{lagDS1}) and (\ref{lagDS2}). We remind that in this counting
$F_0^2$ is order $N_c$, while $B_0$ is order 1 \cite{GL1}.
The effective operators in (\ref{weakoper}) are 
four-quark operators and the leading contributions to them
 are of order $N_c^2$ \cite{tHooft}.
In particular the leading large $N_c$ contribution (i.e. in the absence of
 gluonic corrections) 
to the $\vert\Delta S\vert =1$ operator comes from the 
one $W$-exchange diagram, while the Box diagram leads the leading contribution
to the $\vert\Delta S\vert =2$ operator. In both cases 
only one effective four-quark operator arises in the large $N_c$ limit:
\be
{\cal O}_{\Delta S=1}(x) = Q_2 (x)\equiv 
4 \left(\overline{s}_L \gamma^\mu
u_L \right)(x) \left(\overline{u}_L \gamma_\mu d_L \right) (x)\, ,
\ee
for $\vert\Delta S\vert =1$ transitions and 
\be
{\cal O}_{\Delta S=2}(x) = Q_{\Delta S= 2} (x)\equiv 
4 \left(\overline{s}_L \gamma^\mu
d_L \right)(x) \left(\overline{s}_L \gamma_\mu d_L \right) (x)\, ,
\ee
for $\vert\Delta S\vert =2$ transitions. 
Above we defined  $q_L(x) \equiv [(1-\gamma_5)/2]\, q(x)$ and summation
over colour indices is understood inside each  bracket.
For $N_c\to\infty$ both currents bosonize independently.
So the terms in (\ref{lagDS1}) and (\ref{lagDS2})
with two flavour traces, each of the current type,
may receive contributions of order $N_c^2$.
However, to assign the correct $1/N_c$ counting 
to single and double flavour traces in the weak effective Lagrangian the
 operators have to be traceless.
This can be best seen at the quark level. Light four-quark 
operators are described by SU(3)$_L$ $\times$ SU(3)$_R$ tensors 
\be
\hat{t}^{ij,kl} (\bar{q}_{i} \Gamma q_k)
(\bar{q}_{j} \Gamma'  q_l) 
\ee
with  $\Gamma^{(')}$ the adequate  Dirac structure (pseudo-scalar, scalar,
axial-vector and vector)\footnote{The tensor $\hat{t}$ can  always be decomposed
in a symmetric part $\hat{t}^{ij,kl} = \hat{t}^{ji,lk}$
and an antisymmetric part $\hat{t}^{ij,kl} = -\hat{t}^{ji,lk}$.
We assume $\hat{t}$ is symmetric. If $\hat t$ is antisymmetric
then $\hat t$ is flavour traceless and the couplings modulating it
are order $N_c^2$.}.
  The coupling modulating the 
traceless part of $\hat{t}$, namely $\hat{t}^{ij,kl} - 
(2/3) \, \hat{t}^{mj,ml}$ is leading in the $1/N_c$ counting, i.e. of 
order $N_c^2$. The rest of $\hat t$ is of order $N_c$
since it has an additional flavour trace which is $1/N_c$ suppressed.
Already at order $p^2$ the weak octet Lagrangian, if not written in 
terms of flavour traceless operators, does contain double trace terms 
which are of leading order $N_c^2$.  
At order $p^4$, the basis of 27-plet operators in (\ref{27oper}), the one
in (\ref{S=2oper}) and the one used in Kambor, Missimer, and Wyler in 
\cite{KMW1} are written in terms of traceless operators so that
 the $1/N_c$ counting is correct in those cases. 

Neither the octet basis we use in (\ref{8oper}) nor
the one in \cite{KMW1} are written in terms of traceless operators
so that one has to proceed in two steps to do the correct $1/N_c$ counting.
First, writing the octet Lagrangian in terms of flavour traceless
operators. The second step eventually needed is the reduction to a 
minimal basis, which involves the use of the eqs. of motion, integration 
by parts and Cayley-Hamilton relations. Again, some of these relations can 
spoil the correct $1/N_c$ counting.
In this respect the counting given in \cite{KMW1} for the octet sector is 
not fully correct, while some of the Cayley-Hamilton relations used
in  \cite{EF, EKW} for the reduction to a minimal basis were not appropriate 
for a correct large $N_c$ counting.

A way of deducing the leading in $1/N_c$ contributions to the weak couplings
 at order $p^4$ is the use of the strong factorization assumption
(See \cite{EKW} and references therein)
where all the weak parameters in (\ref{lagDS1}) are known to all
orders in CHPT, in terms of the parameters of the strong effective 
Lagrangian. All the large $N_c$ contributions to the weak parameters are
 contained in their factorizable part. 
Up to order $p^4$ (and assuming the fudge factor of naive factorization
 $k_f =1$)
 the weak couplings receive the following large $N_c$ contributions 
\ba
\label{leadingNc}
G_8&=&1\, , \quad G_{27}= 1 \, , \nonumber \\
E_{10}&=& 8 L_4 + 2 L_5\, , \quad E_{11}=
8 L_4 + 4 L_5 \, , \quad E_{12}= -4 L_4 - \frac{4}{3} L_5\, , \quad
E_{13}=-4 L_4 \, , \nonumber \\
D_4 &=& 4 L_5 \, , \quad D_7 = 8 L_4 \,  
\ea
and all the others in (\ref{lagDS1}) being zero.
The strict large $N_c$ limit would also imply $L_4=0$. 
The factorization assumption actually 
corresponds to keeping $L_4$ non-zero in (\ref{leadingNc})
and leaving $G_8$ and $G_{27}$ as free parameters. 

At next-to-leading order in $1/N_c$ (i.e. in the presence of gluonic
 corrections) all the weak parameters receive corrections both 
non-factorizable and factorizable of  short- and long-distance origin.
$G_8 \neq G_{27}$ at this order. Analogously all the $p^4$
couplings in (\ref{leadingNc}) and the rest of the counterterms in 
(\ref{lagDS1}) which are zero in the large $N_c$ limit do get unknown 
contributions at next-to-leading order. Their estimate is one of the main
 challenges of low-energy physics. 
Above we have included the factorizable next-to-leading in $1/N_c$ 
contributions too -- the $L_4$ parts.
As shown in Table \ref{table1} the octet counterterms $E_{10}, E_{11}, 
E_{12}$ and the 27-plet counterterm $D_4$ get leading order contributions in 
$1/N_c$.
To arrive at the basis we are using in (\ref{lagDS1}) 
and (\ref{lagDS2}) we made use of the equations of motion,
 partial integrations and Cayley-Hamilton relations.
The latter have been used in such a way that the $1/N_c$
counting is not broken. The use of the equations of motion
does not break the counting either. However
some integrations by parts do, e.g. one can remove the
apparent current--current structure of some operators in this way.
Therefore we have to do the 
$1/N_c$ counting of the other counterterms in Table \ref{table1} 
before the integrations by parts are done. Afterwards, the counting is 
translated to the basis in (\ref{lagDS1})\footnote{In the strict
$N_c\to\infty$ limit the singlet $\eta_1$ degree of freedom needs to be
included. This one should afterwards be integrated out, leading to
counterintuitive $N_c$-counting as is the case for $L_7$ in the strong
sector \cite{GL1}.
Since the numerical value of $L_7$ is such that this counting
seems inappropriate for our real world we neglect this issue. It should
however be included in modeling approaches since there can be sizeable 
contributions from it as e.g. seen in $B_K$ \cite{BPBK}.}.

We summarize  the
$1/N_c$ counting of the weak couplings we are using.
 The ${\cal O}(p^2)$ couplings $G_8$ and $G_{27}$
are order 1 while $G_8'$ is order $1/N_c$ (notice that $F_0^4$ has been
 factored out).
The ${\cal O} (p^4)$ couplings $E_{10}$, $E_{11}$, $E_{12}$, 
and $D_4$ are order $N_c$ ($F_0^2$ has been factored out in (\ref{lagDS1})). 
The couplings $E_1$, $E_3$, $E_4$, $E_5$, $E_{13}$,
$E_{15}$, $D_1$, $D_2$, $D_5$, $D_6$, and $D_7$
are order 1. The combination of couplings
$E_2+ 2 E_1 /3$ is order $1/N_c$. 

\section{Two-Point Functions}
\label{sectwopoint}
\setcounter{equation}{0}

In this section we give the two-point functions
in (\ref{def2point}) at order $p^4$ for all
the relevant $ij$ combinations.

\subsection{Strangeness Zero}

\begin{figure}
\begin{center}
\leavevmode\epsfxsize=12cm\epsfbox{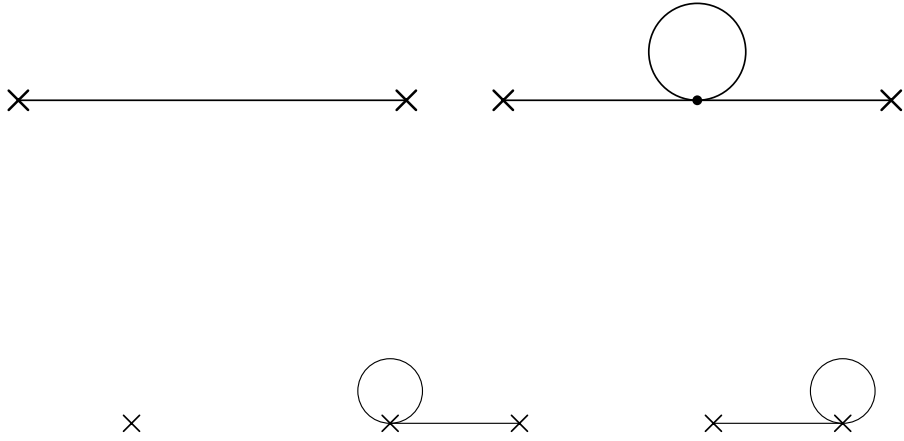}
\end{center}
\caption{The diagrams contributing to the $\Delta S=0$ two-point functions.
A line is a pseudoscalar meson propagator, a dot a strong vertex with only
meson legs and a cross a vertex from the strong Lagrangian with one or more
insertions of the external pseudoscalar currents.
\label{figure1}}
\end{figure}

Here we give the two-point Green's functions 
in (\ref{def2point}) with $i=j$ to order $p^4$.
Notice that those conserving strangeness with $i\ne j$
 vanish since $m_u=m_d$.
The poles of these two-point functions 
define the masses of the corresponding mesons 
and set the renormalization factors $Z_i$ for the pseudoscalar sources $P^i$
needed for the reduction procedure.

\be
\Pi_{ii}(q^2) \equiv - \left[\frac{ Z_{i}}{{q^2-m_{i}^2}} + Z_i'\right]\, .
\ee
To order $p^4$, using $m_u=m_d=\overline m$ and neglecting 
electromagnetic corrections, we get from the diagrams in Fig. \ref{figure1}
\ba
Z_{\pi^0}=Z_{\pi^+}& =& 2 B_0^2 F_0^2 \left[ 
1+ \frac{8}{F_0^2} ( 2 m_K^2 + m_\pi^2) \, 
(4 L_6-L_4) + \frac{8}{F_0^2} m_{\pi}^2 (4 L_8 - L_5) 
\right. \nonumber \\ &-& \left. 
2  \mu_\pi - 2\mu_K - \frac{2}{3}
\mu_{\eta_8} \right] \, ; \nonumber \\
Z_{K^0}=Z_{K^+} &=& 2 B_0^2 F_0^2 \left[ 
1+ \frac{8}{F_0^2} ( 2 m_K^2 + m_\pi^2) \, 
(4 L_6-L_4) + \frac{8}{F_0^2} m_K^2 (4 L_8 - L_5) 
\right. \nonumber \\ &-& \left. 
\frac{3}{2} \mu_\pi - 3 \mu_K - \frac{1}{6} 
\mu_{\eta_8}  \right] \, ; \nonumber \\
Z_{\eta_8}&=&2 B_0^2 F_0^2 \left[ 
1+ \frac{8}{F_0^2} ( 2 m_K^2 + m_\pi^2) \, 
(4 L_6-L_4) + \frac{8}{F_0^2} m_{\eta_8}^2 (4 L_8 - L_5) 
\right. \nonumber \\
&-& \left.  2 \mu_\pi - \frac{2}{3} \mu_K - 
2 \mu_{\eta_8} \right] \, .
\ea
\ba
Z_{\pi^0}'=Z_{\pi^+}'=Z_{K^0}'=Z_{K^+}'=
Z_{\eta_8}'= 8 B_0^2 (2 L_8-H_2) \,,
\ea
and
\ba
m_{\pi^0}^2&=&m_{\pi^+}^2= 2 \overline m B_0 \left[
1 + \frac{8}{F_0^2} ( 2 m_K^2 + m_\pi^2) (2 L_6-L_4)
\right. \nonumber \\
&+& \left.\frac{8}{F_0^2} m_\pi^2 (2L_8-L_5)
+ \mu_\pi - \frac{1}{3} \mu_{\eta_8} \right] \, ; \nonumber \\ 
m_{K^0}^2&=&m_{K^+}^2= (\overline m + m_s) B_0 \left[
1 + \frac{8}{F_0^2} ( 2 m_K^2 + m_\pi^2) (2 L_6-L_4)
\right. \nonumber \\
&+& \left. \frac{8}{F_0^2} m_K^2 (2L_8-L_5)
+ \frac{2}{3} \mu_{\eta_8} \right] \, ; \nonumber \\ 
m_{\eta_8}^2&=& \frac{2}{3} (\overline m + 2 m_s) B_0 \left[
1 +  \frac{8}{F_0^2} ( 2 m_K^2 + m_\pi^2) (2 L_6-L_4)
\right. \nonumber \\
&+& \left. \frac{8}{F_0^2} m_{\eta_8}^2 (2L_8-L_5)
+ 2 \mu_K  -\frac{4}{3} \mu_{\eta_8} \right] \nonumber \\ 
&+& 2 \overline m B_0 \left[ -\mu_\pi + \frac{2}{3} \mu_K
+ \frac{1}{3} \mu_{\eta_8} \right] + B_0^2 (m_s - \overline m)^2 
\frac{128}{9\, F_0^2} \, (3 L_7 + L_8)  \nonumber \, . \\
\ea
For completeness and later use we also quote
the decay constants $f_\pi$, $f_K$, and $f_{\eta_8}$, to
the same order:
\ba
f_{\pi^0}^2=f_{\pi^+}^2&=& F_0^2 \left[ 1 + 
\frac{8}{F_0^2} (2 m_K^2 + m_\pi^2) L_4 +
\frac{8}{F_0^2} m_\pi^2 L_5  - 4 \mu_\pi - 2 \mu_K 
\right] \nonumber \, ; \\
f_{K^0}^2=f_{K^+}^2&=& F_0^2 \left[ 1 + 
\frac{8}{F_0^2} (2 m_K^2 + m_\pi^2) L_4 +
\frac{8}{F_0^2} m_K^2 L_5  - \frac{3}{2} \mu_\pi - 3 \mu_K 
- \frac{3}{2} \mu_{\eta_8} \right] \nonumber \, ; \\
f_{\eta_8}^2&=& F_0^2 \left[ 1 + 
\frac{8}{F_0^2} (2 m_K^2 + m_\pi^2) L_4 +
\frac{8}{F_0^2} m_{\eta_8}^2 L_5  - 6 \mu_K 
\right] \nonumber \, ; \\
\ea 
We use the notation \cite{GL1}
\be
\mu_i \equiv \frac{m_i^2}{32 \pi^2 F_0^2} \ln\left(\frac{m_i^2}{\nu^2}
\right) \, . 
\ee

\subsection{Strangeness One}

\begin{figure}
\begin{center}
\leavevmode\epsfxsize=12cm\epsfbox{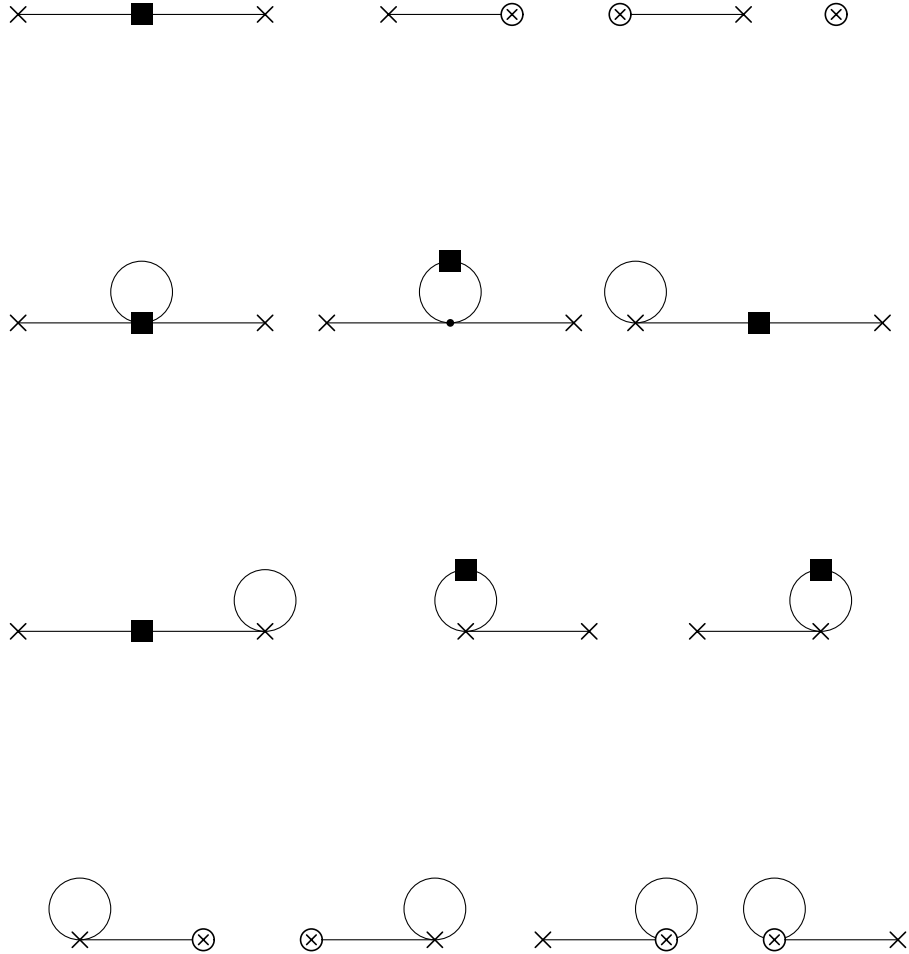}
\end{center}
\caption{The diagrams contributing to the $\vert\Delta S\vert =1$
and $\vert\Delta S\vert =2$ two-point
functions up to one loop. In addition to the symbols of Fig. \ref{figure1},
the full square is a weak vertex with only meson legs and the circled cross
is a weak vertex with one or more insertions of the external pseudoscalar
currents.
\label{figure2}}
\end{figure}

In this section we give the renormalized two-point Green's functions 
in (\ref{def2point}) up
to order $p^4$. We define them as 
\be
\Pi_{ij}(q^2) \equiv
\Pi_{ij}(q^2)\left|_{\mbox{\tiny{Count}}} \right. 
+ \Pi_{ij}(q^2) \left|_{\mbox{\tiny{Logs}}} \right. 
\ee
Only the analytic contributions from the counterterm Lagrangian
 $\Pi_{ij}(q^2)\left|_{\mbox{\tiny{Count}}}\right.$ 
are written here, while we give the non-analytic contributions from the one 
loop integration  $ \Pi_{ij}(q^2) \left|_{\mbox{\tiny{Logs}}}\right. $
 in  Appendix \ref{applogs1}.
The contributing diagrams are shown in Fig. \ref{figure2}
With the pseudo-scalar sources  $P^i(x) = P^{K^0}(x)$
and $P^j(x)=P^{\eta_8}(x)$ the two-point Green's function is given by
\ba
 \Pi_{K^0\eta_8}(q^2) \left|_{\mbox{\tiny{Count}}}  \right. &=&
- \frac{\sqrt{Z_{K^0} Z_{\eta_8}}}
{(q^2-m_{K^0}^2)(q^2-m_{\eta_8}^2)} \, 
\frac{C}{\sqrt 6}  \, \nonumber \\
& \times & \, 2 \, \frac{F_0^4}{f_{K^0} f_{\eta_8}}  
\left[ q^2 (G_{27}-G_8 + \frac{f_{\eta_8}^2 + f_K^2}{F_0^2} G_8') 
- m_{\eta_8}^2
\frac{f_{\eta_8}^2}{F_0^2}\, G_8'  
\right. \nonumber \\
&+& \frac{2}{F_0^2}  \left\{ 
2 q^4 \left( G_8 E_3 - G_{27}  D_2 \right) 
\right. \nonumber \\
&+&  q^2 m_K^2 \left[ G_8 \left( 4 E_1 + 4 E_2 + 
\frac{16}{3} E_3 + 8 E_4
   + 6 E_5 \right. \right. \nonumber \\ &-&
\left.  2 E_{10} - 4 E_{11} - 8 E_{12} \right) \nonumber \\ &-&
\left.  \frac{G_{27}}{3} \left(4 D_1 - 7 D_4 - D_6  -6 D_7 \right)
- 8 G_8' \left( 4 L_6 + \frac{7}{3} L_8 \right) \right]  \nonumber \\ 
&+& q^2 m_\pi^2 \left[ G_8 \left( 2 E_2 -\frac{16}{3} E_3 -8 E_4 
- 6 E_5 + 3 E_{11} + 8 E_{12} \right) 
\right. \nonumber \\  &+& \left. 
\frac{G_{27}}{3} \left( 4 D_1 - D_4  - D_6 +3 D_7 \right) 
- 8 G_8' \left( 2 L_6 - \frac{1}{3} L_8 \right) \right]
\nonumber \\
&-& \frac{8}{3} m_K^4 \left[ G_8 \left( E_1 + E_2 \right) 
- \frac{G_{27}}{3}
 D_1 -  8 G_8' \left( L_6 + \frac{2}{3} L_8 \right) \right]
\nonumber \\
&+& \frac{2}{3} m_K^2 m_\pi^2 \left[ G_8 
\left( E_1 - E_2 \right)  -  5 \frac{G_{27}}{3} D_1 
+ 8 G_8' \left( L_6 - \frac{4}{3} L_8 \right) \right] \nonumber\\
&+& \left. \left. \frac{1}{3} m_\pi^4
\left[ G_8 E_2 + 2 \frac{G_{27}}{3} D_1  
- 8 G_8' \left( L_6 - \frac{1}{3} L_8 \right) \right] \right\} \right] \, .
\ea
Notice that the renormalized meson masses $m_\pi , m_K, m_{\eta_8}$ are used
 everywhere. For the case of pseudo-scalar sources $P^i(x)=P^{K^0}(x)$ and
$P^j(x)=P^{\pi^0}(x)$ the result is
\ba
\Pi_{K^0\pi^0}(q^2) \left|_{\mbox{\tiny{Count}}}\right.  &=&
-\frac{\sqrt{Z_{K^0} Z_{\pi^0}}}
{(q^2-m_{K^0}^2)(q^2-m_{\pi^0}^2)} \, \frac{C}{\sqrt 2}
\, \nonumber \\ &\times&
\, 2 \frac{F_0^4}{f_{K^0} f_{\pi^0}} \, 
\left[ q^2 \left(G_{27}-G_8+ \frac{f_\pi^2+f_K^2}{F_0^2} G_8'\right) 
- m_{\pi}^2 \frac{f_\pi^2}{F_0^2} G_8'
\nonumber \right. \\
&+& \frac{2}{F_0^2} \left\{ 2 q^4 \left( G_8 E_3 - G_{27} D_2 \right) 
\right. \nonumber \\ &+&
q^2 m_K^2 \left[ G_8 \left( 4 E_1 + 4 E_2 - 2 E_5 - 2 E_{10} \right)
\right. \nonumber \\
&-& \left. \frac{G_{27}}{3} \left( 4 D_1 - 3 D_4 - D_6 - 6 D_7 \right) 
- 8 G_8' \left( 4 L_6 + L_8 \right) \right] \nonumber \\
&+& q^2 m_\pi^2 \left [ G_8 \left( 2 E_2 + 2 E_5 - E_{11} \right)
\right. \nonumber \\
&+& \left. 
 \frac{G_{27}}{3} \left(4 D_1 + 3 D_4 - D_6 + 3 D_7   \right) 
- 8 G_8' \left( 2 L_6 + L_8 \right)  \right]
\nonumber \\
&-& 2 m_K^2 m_\pi^2 \left[ G_8 \left( E_1 + E_2\right) - \frac{G_{27}}{3}
 D_1 - 8 G_8' L_6  \right] \nonumber \\
&-& \left. \left. m_\pi^4 \left[ G_8 E_2 + 2 \frac{G_{27}}{3} D_1 
-8 G_8' \left( L_6 + L_8 \right) \right]  \right\} \right] \, .
\ea

Finally, for  $P^i(x)=P^{K^+}$
and $P^j(x)=P^{\pi^+}$ we obtain 
\ba
\Pi_{K^+\pi^+}(q^2) \left|_{\mbox{\tiny{Count}}}\right.  &=&
-\frac{\sqrt{Z_{K^+} Z_{\pi^+}}}
{(q^2-m_{K^+}^2)(q^2-m_{\pi^+}^2)} \, C
\, \nonumber \\ &\times&
\, 2 \frac{F_0^4}{f_{K^+} f_{\pi^+}} \,
\left[ q^2 \left(G_8 + 2 \frac{G_{27}}{3} - \frac{f_\pi^2+f_K^2}{F_0^2} 
G_8'\right) + m_{\pi}^2  \frac{f_\pi^2}{F_0^2} G_8'
\nonumber \right. \\
&+& \frac{2}{F_0^2} \left\{-2 q^4 \left( G_8 E_3 + 
2 \frac{G_{27}}{3} D_2
\right) \right. \nonumber \\
&-& q^2 m_K^2 \left [ G_8 \left( 4 E_1 + 4 E_2 -  2 E_5 -2 E_{10}
\right) \right. \nonumber \\
&+& \left. \frac{G_{27}}{3} \left( -4 D_1 - 2 D_4 + D_6
 - 4 D_7  \right)  - 8 G_8' \left( 4 L_6 + L_8 \right)  \right] \nonumber \\
&-& q^2 m_\pi^2 \left [ G_8 \left( 2 E_2 + 2 E_5 - E_{11} 
\right) \right.
\nonumber \\ 
&+& \left. \frac{G_{27}}{3} \left( 4 D_1 - 2 D_4  - D_6 
- 2 D_7 \right) -8 G_8' \left( 2 L_6 + L_8 \right) \right] \nonumber \\
&+& 2 m_K^2 m_\pi^2 \left[ G_8 \left( E_1 + E_2 \right)  
- \frac{G_{27}}{3} D_1 - 8 G_8' L_6 \right] \nonumber \\
&+& \left. \left. m_\pi^4 \left[ G_8 E_2 + 2 \frac{G_{27}}{3} D_1 
-8 G_8' (L_6+L_8) \right] \right\} \right]\, .
\ea

\subsection{Strangeness Two}

The two-point Green's function for the $\Delta S=2$ transition
was already calculated in \cite{BPBK}. We include it here for sake of 
completeness. With the notation used in the present work we need
the $\Delta S=2$ part of the two-point Green's function 
in (\ref{def2point}) with $P^i(x)=P^{K^0}$ and 
$P^j(x)=P^{\overline K^0}$. This gives
\ba
\Pi_{K^0  \overline K^0}^{\Delta S=2}(q^2) 
\left|_{\mbox{\tiny{Count}}}\right.  &=&
-\frac{Z_{K^0}}{(q^2-m_{K^0}^2)^2} \, C_{\Delta S=2}
\, G_{27} \, 4 \frac{F_0^4}{f_{K^0}^2} \,
\left[ q^2  + \frac{1}{F_0^2} \left\{- 4 q^4 D_2 
\right. \right. \nonumber \\ &+&  
\left. \left. 2 q^2 \left(2 D_4 m_K^2 + D_7  
\left( 2 m_K^2 + m_\pi^2 \right) \right) -
4 D_1 \left(m_K^2-m_\pi^2\right)^2 
\right\}   \right] \, . \nonumber \\ 
\ea
The non-analytic contributions are in Appendix \ref{applogs1}.
There is another contribution $(\Delta S=\pm 1)^2$
to this two-point function which comes
from expanding the exponential in (\ref{def2point}) up to second order.
They are the so-called long-distance contributions to
$K^0-\overline K^0$ mixing.  

\section {$K \to \pi \pi$ Amplitudes}
\setcounter{equation}{0}

\begin{figure}
\begin{center}
\leavevmode\epsfxsize=12cm\epsfbox{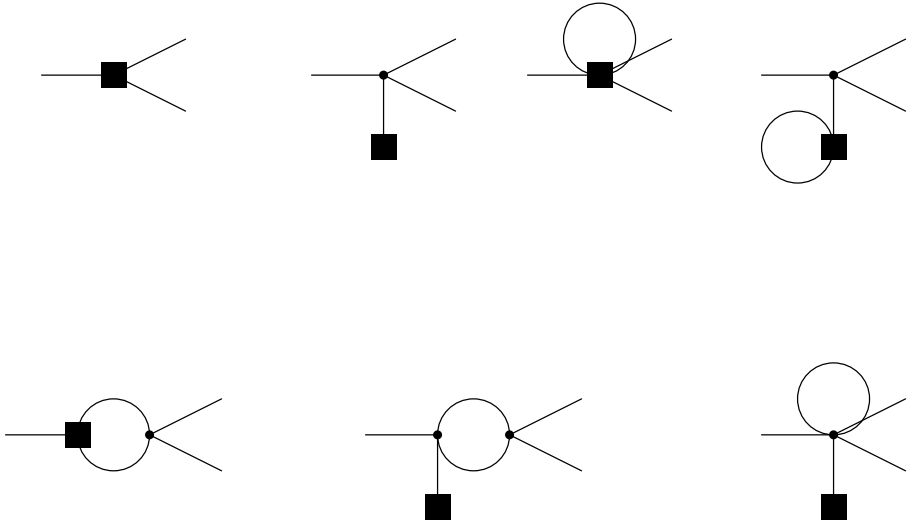}
\end{center}
\caption{The diagrams contributing to on-shell $K\to\pi\pi$ amplitude up to
  one loop. Symbols as in Figs.
\ref{figure1} and \ref{figure2}. In addition, renormalization of
 external legs and internal propagators
of the tree level diagrams have to be added.
\label{figure3}}
\end{figure}

We have the following decomposition
into definite isospin quantum numbers invariant amplitudes 
[$A\equiv -i T$],
\ba
A \left[ K_S \to \pi^0 \pi^0 \right] \equiv
\sqrt{\frac{2}{3}}A_0 - \frac{2}{\sqrt 3} A_2 \, ; \nonumber \\
A \left[ K_S \to \pi^+ \pi^-\right] \equiv
\sqrt{\frac{2}{3}}A_0 + \frac{1}{\sqrt 3} A_2 \, ; \nonumber \\
A \left[ K^+ \to \pi^+ \pi^0\right] \equiv
\frac{\sqrt 3}{2} \, A_2 \, .
\ea
Where $K_S \simeq K^0_1 + \varepsilon \, K^0_2$, 
$K^0_{1(2)} \equiv (K^0-(+)\overline{K^0})/\sqrt 2$, 
and CP $K^0_{1(2)} = +(-)K^0_{1(2)}$.
Since CP violation is small we set $\varepsilon =0$
and therefore $\Im m \,  G_8 =0$, 
$\Im m \, G_{27} =0$, and $\Im m \, G_8' =0$.
We have also included the final state interaction phases into the
amplitudes $A_0$ and $A_2$. For the isospin 1/2 amplitude we have
\ba
A_0 \equiv -i a_0 \, e^{i \delta_0} \, 
\ea
and for the isospin 3/2 amplitude we have
\ba
A_2 \equiv - i a_2 \, e^{i \delta_2} \, .
\ea 
To order $p^2$ we get 
\ba
a_0&\equiv& a_0^8+ a_0^{27}= C \left[ G_8 + \frac{1}{9} G_{27} \right]
\, \sqrt{6} \, F_0\, (m_K^2-m_\pi^2) \,  , \nonumber \\
a_2&=&C \, G_{27} 
\, \frac{10 \sqrt 3}{9} \, F_0\, (m_K^2-m_\pi^2) \,  , 
\ea
and 
\ba
\delta_0 = \delta_2 = 0 \, . 
\ea
The order $p^4$ counterterms contributions 
to $A_0^8$, $A_0^{27}$, and 
$A_2$ (see Appendix \ref{applogs2} for the non-analytic contributions) are
\ba
\Im m \, A_0^8 \left|_{\rm Count.}\right.  &=&
-C \, G_8 \sqrt{6} \, \frac{F_0^4}{f_K f_\pi^2} \, 
(m_K^2-m_\pi^2) \nonumber \\
&\times& \left[ 1 +
\frac{2}{F_0^2} \left[ m_\pi^2 \left( -2 E_1 - 4 E_2 - 
2  E_3 +2  E_{10} +  E_{11} + 4  E_{13} \right)
\right. \right. \nonumber \\  
&+& \left. \left. m_K^2 \left(   E_{10} - 2  E_{13} +  E_{15} \right) 
\right] \right] \nonumber \\ 
&-&  C \, G_8' 8 \sqrt{6} \, \frac{F_0^2}{f_K f_\pi^2} \, 
(m_K^2-m_\pi^2) \nonumber \\ &\times&
 \left[ m_\pi^2 \left(-4 L_4-L_5+8 L_6 +4 L_8\right)
+ 2 m_K^2 L_4    \right]   \nonumber \\ 
\ea
and
\ba
\Im m \, A_0^{27} \left|_{\rm Count.}\right.  &=&
-C \, G_{27} \frac{\sqrt{6}}{9} \, 
\frac{F_0^4}{f_K f_\pi^2} \,  (m_K^2-m_\pi^2) 
\nonumber \\ &\times& 
\left[ 1 + \frac{1}{F_0^2} \left[
2 m_\pi^2 \left( -6 D_1 - 2 D_2 + 2 D_4 + 6 D_6 + D_7\right) 
\right. \right. \nonumber \\ &+& \left. \left. 
 m_K^2 \left( D_4 - D_5 -9 D_6 + 4 D_7 \right) \right] \right]
\nonumber \\
\ea
and
\ba
\Im m \, A_2 \left|_{\rm Count.}\right.  &=&
- C \, G_{27} \frac{10 \sqrt 3}{9} \, 
\frac{F_0^4}{f_K f_\pi^2} \,  (m_K^2-m_\pi^2) 
\nonumber \\ &\times& \left[ 1 +
\frac{1}{F_0^2} \left[ 2 m_\pi^2 \left( 
- 2 D_2 + 2 D_4 + D_7 \right)
+ m_K^2 \left( D_4 - D_5 + 4 D_7 \right) \right]
\right] \, . \nonumber \\ 
\ea
The diagrams are depicted in Fig. \ref{figure3}. In addition there are
the corrections on the external legs and on the internal propagators
of the tree level diagrams.

\section {$K \to \pi \pi$ from $K \to \pi , \eta_8$ Amplitudes}
\label{kptokpp}
\setcounter{equation}{0}

We discuss here the information we can extract from the $K\to \pi ,\eta_8$ 
two-point functions
and make some remarks about the parameters needed for $K\to\pi\pi$ we
cannot obtain. As discussed in Sect. \ref{weakmass}, the weak
mass term contribution to $K\to\pi\pi$ decays can 
be absorbed in a redefinition of the other
coefficients \cite{KMW1}, while this is 
not true in the case of the two-point functions. 
In typical approaches used in lattice QCD or effective models like 
the one proposed in \cite{Antonelli},
the weak mass term should be treated as an extra 
parameter in the determination of $K \to \pi \pi$ amplitudes at order 
$p^4$.

A few other remarks are needed here. The contribution from the $27_L$ and
$8_L$ cannot be easily disentangled in general, since for $m_s\ne m_d=m_u$
 the two components are mixed by higher order effects in $m_s-\overline{m}$.
The $\Delta I = 1/2$ and $\Delta I=3/2$ contributions are however separate.
The two-point function $\Pi_{K^0\eta_8}(q^2)$ and
the combination $(1/\sqrt 2) \Pi_{K^0\pi^0}(q^2) - \Pi_{K^+\pi^+}(q^2)$
are pure $\Delta I=1/2$, while 
$\sqrt 2 \Pi_{K^0\pi^0}(q^2) + \Pi_{K^+\pi^+}(q^2)$ is pure $\Delta I=3/2$.

The $K \to \pi$ and $K \to \eta_8$ two-point functions defined here
are not measurable in experiments, so obtaining
the counterterms from them implies
that one has to calculate the relevant two-point functions
either using lattice QCD \cite{Sharpe, Martinelli2} or
using other hadronic approaches \cite{BPBK,BPdashen}.

Recent work on  $\Delta I = 3/2$ transitions in quenched 
CHPT \cite{quenched3/2} gives numerically consistent
results with quenched results on $B_K$. 
The main uncertainty are the unknown quenched
counterterms: $G_{27}$ and
$D_i$'s. The value of several of them can be similarly extracted 
from two-point functions but this has not been done so far \cite{aoki}.
 In the near future it might be possible to 
calculate the two-point functions 
we propose in (\ref{def2point})
both in the quenched and unquenched case and the $K\to\pi\pi$ amplitude
only quenched. The comparison of all the constants that can be calculated
using the simpler two-point correlator would then be a useful tool to estimate
quenching errors on the remainder.

We now discuss the expressions of Sect. \ref{sectwopoint} to check which
constants are obtainable.
The order $p^4$ counterterms $E_{13}$, $E_{15}$, and $D_5$
cannot be obtained from the $\Pi_{ij}(q^2)$ since they do not contribute
to them. They do however contribute to $K \to \pi \pi$. 
Their value in the large $N_c$ limit 
and with the factorization assumption is known, see Sect. \ref{Nccounting}.

In the chiral limit we can get $G_8$, $G_{27}$, $E_3$, and $D_2$. Away from
 the chiral limit, we can get $G_8'$ from the terms
quadratic in the meson masses. From the terms quartic
in the meson masses we can get $E_1$, $E_2$, and $D_1$. 
The terms proportional to $q^2 m_{\pi,K}^2$ allow us to obtain
the combination $2E_{10}+E_{11}$, $D_4$, $D_6$, and $D_7$.
The latter also determine two more combinations 
of couplings, though not needed for $K \to \pi \pi$: $E_4+E_{12}$
and $E_5+E_{10}$. $E_{10}$ is needed for $K\to\pi\pi$ but cannot be
separately disentangled from the two-point functions $\Pi_{ij}(q^2)$.

We have determined three order $p^2$ couplings and
eleven order $p^4$ ones. We have in addition four
relations which are independent of the value of
the couplings to test chiral symmetry at order $p^4$.
This is the main result of this manuscript.
We are left with four unknowns $E_{10}$, $E_{13}$, $E_{15}$, and $D_5$
in order to extrapolate to $K\to\pi\pi$.

What can be said about the missing coefficients. Three kind of arguments 
can be used:\\[0.2cm]
{\bf 1) Order of magnitude:}
We know the leading in $1/N_c$ contributions
to all of them, Eq. (\ref{leadingNc}). In particular, 
$E_{13}$, $E_{15}$, and $D_5$
only receive non-factorizable contributions. 
{}From the discussion above, we have seen that we can  determine 
eleven next-to-leading in $1/N_c$ contributions to the
couplings, this should give us some information on 
the ones we cannot get. 

For instance, assuming that
all the $1/N_c$ contributions are of the same order, 
since $2E_{10} + E_{11}$ is obtained from terms of the
 type $q^2 m_{\pi,K}^2$ and we know its factorizable 
contribution
\be
(2 E_{10}+E_{11})_{\rm Factorizable} =8 L_5 + 24 L_4 \, , 
\ee
we can use (conservative choice) the next-to-leading 
result we have for $2E_{10}+E_{11}$
as a good estimate for $E_{13}$, $E_{15}$ and the non-factorizable
part of $E_{10}$. Analogously for $D_5$ we can use the
result
\be
(D_4)_{\rm Factorizable}=4 L_5 \, , 
\quad (D_7)_{\rm Factorizable}
= 8 L_4 \, ,
\ee
for predicting the non-factorizable part of $D_5$.
\\[0.2cm]
{\bf 2) Resonance Saturation:} Another possibility is using estimates
of higher order parameters coming from resonance exchange saturation, 
as done in \cite{EKW}. It is well known and experimentally proven that
vector and/or axial-vector dominance is not at work in the weak sector (which
is instead the case in the strong one). Within the tensor formulation of
vector (axial-vector) resonances used in \cite{EKW} all the counterterms 
in (\ref{lagDS1})
 only receive contributions from scalar and/or pseudo-scalar resonances.
In Table \ref{tableres} we summarize the resonance contributions to the
 counterterms contained in (\ref{lagDS1}). 
The notation is the one used in \cite{EKW}.
For the 27-plet the derivation is done in Appendix \ref{appres}.
\begin{table}[htb]
\begin{center}
\begin{tabular}{|c|c|c|c|}
\hline
   & $S$ & $S_1$ & $P$ \\
\hline 
&&&\\
$E_1$ & $2c_m g_S^1$ & -- & --  \\
$E_2$ & $c_m \left (g_S^2-\frac{\dis 2}{\dis 3}g_S^1\right )$ & 
$\tilde{c}_m\tilde{g}_S^1$ & -- \\
$E_3$ &  -- & -- & $2 d_m g_P^1$ \\
$E_4$ &  -- & -- &{\small $ d_m \left 
( -\frac{\dis  2g_P^1}{\dis 3}+g_P^2\right )$} \\
$E_5$ & $c_m g_S^3$ & -- & $-d_m g_P^3$ \\
$E_{10}$ & $c_d(g_S^1+g_S^2)+\frac{\dis c_m}{\dis 3} g_S^4$ 
& $\tilde{c}_m\tilde{g}_S^2 $
 & -- \\
$E_{11}$ & $c_d g_S^2-\frac{\dis 2}{\dis 3}c_m g_S^4$ & 
$\tilde{c}_m\tilde{g}_S^2 $
 & -- \\
$E_{12}$ & $-\frac{\dis c_d}{\dis 2}g_S^2
+c_m\left (\frac{\dis g_S^4}{\dis 3}+\frac{\dis g_S^5}{\dis 2}\right )$ & 
$-\frac{\dis \tilde{c}_m}{\dis 2}\tilde{g}_S^2 $& --  \\
$E_{13}$ &{\small $-c_d\left ( \frac{\dis 2g_S^1}{\dis 3}+
\frac{\dis g_S^2}{\dis 2}\right ) 
+c_m \left ( g_S^6+\frac{\dis g_S^4}{\dis 3}\right )$}&
{\small $\tilde{c}_d\tilde{g}_S^1
-\frac{\dis \tilde{c}_m}{\dis 2}\tilde{g}_S^2$}& -- \\
$E_{15}$ & $-c_d g_S^3$ & -- & $-d_m g_P^4$ \\
$D_1$ & $ c_m\bar{g}_S^1$ & -- & --  \\
$D_2$ & -- & -- & $ -d_m\bar{g}_P^1$  \\
$D_4$ & $ c_m\bar{g}_S^2$ & -- & --  \\
$D_5$ & -- & -- & $ -d_m\bar{g}_P^2$  \\
$D_6$ & $c_d\bar{g}_S^1+c_m\bar{g}_S^3 $ & -- & --  \\
$D_7$ & -- & $ \tilde{c}_m\tilde{\bar{g}}_S^1$ & --   \\
&&&\\
\hline
\end{tabular}
\end{center}
\normalsize
\protect\caption{ The contributions to the octet $E_i$ and 27-plet $D_i$ 
counterterms in \protect(\ref{lagDS1}) from scalar octet (S), 
scalar singlet ($S_1$) and pseudo-scalar octet (P) resonance. 
The pseudo-scalar singlet 
$(P_1)$ only contributes to $E_4$ with a term $\tilde{d}_m\tilde{g}_P^1$.
A factor $1/M_R^2$ is pulled out. Vector and axial-vector resonances
exchange do not contribute \protect\cite{EKW}.}  
\label{tableres}
\end{table}
All of the unknown terms $E_{10}, E_{13}, E_{15}$ and $D_5$ only receive
scalar and/or pseudo-scalar contributions. This is consistent with the
consequences of the factorization assumption 
as shown in (\ref{leadingNc}), where the weak counterterms are all expressed
in terms of the strong counterterms $L_4$ and $L_5$, which are in turn
saturated by scalar exchange \cite{EKW,EGPR}. 
A simplified resonance model 
is e.g. the one where pseudo-scalar resonances exchange is neglected
\footnote{One can, of course, use other resonance models
to make the analysis. We mention that in the  vector formulation
with vector fields used in \cite{AP}, not antisymmetric tensor fields
as in \cite{EKW}, $E_{15}$ also receives contribution from vector resonance 
exchange. We do not address here the question of the equivalence of different
resonance models in the weak sector.}. Their contribution is small
in the strong sector \cite{EGPR}. With this reduction several
relations are valid. $E_3=0, E_4=0, D_2=0$ and $D_5=0$ are a test of 
scalar dominance, while
\be
2E^r_{10}+E_{11}^r-\frac{c_d}{c_m}\left(2E_1^r-2E_3^r+3E_2^r\right)=0
\ee
tests the absence of singlet scalar resonance contributions. 
The couplings $c_d$ and $c_m$
are scalar couplings from the strong sector \cite{EKW,EGPR}.
Additional relations are $(c_m/c_d)\;E_{15}^r+E_5^r-E_3^r=0$, valid in the 
absence of pseudo-scalar resonance contributions, and
\be
E_{13}^r = \frac{1}{6}\left[-(2E_{10}^r+E_{11}^r)-
\frac{c_d}{c_m}\left(E_1^r-E_3^r\right)+\frac{c_m}{c_d}
\left(3N_1^r+3N_2^r+6N_3^r\right)\right]\, ,
\ee
valid in the absence of the singlet scalar resonance.
The combination of $N_i$ corresponds to $K_2$ which receives no
vector/axial-vector contributions and
is a combination of the $p^4$ constants that can be determined from $K\to3\pi$
decays \cite{EKW,KDHMW}.
Hence, introducing additional information about $K\to3\pi$ decays
we can use resonance arguments to determine $E_{13}$. 
One main observation is that the direct determination of most of the couplings
from the analysis of the two-point functions offers already a powerful test
of the validity of different resonance saturation assumptions.
\\[0.2cm]
{\bf 3) Factorization:} We can of course also adopt the strong factorization
assumption as often used to get at the undetermined parameters.
Again this procedure can be well tested by the fourteen parameters
we can actually determine and comparing them with the predictions of
Eq. (\ref{leadingNc}) with $G_8$ and $G_{27}$ as free parameters.

\section{The Weak Mass Term  Contributions}
\label{weakmass}
\setcounter{equation}{0}

In the literature there are conflicting opinions about whether
the weak mass term contributes to $K\to\pi$ and $K\to$ vacuum matrix elements.
In \cite{Crewther} the claim is that they do not and 
in \cite{Bernard,Yaouanc,Maiani} they
do. The underlying reason for the difference is that the meson fields
used in those two references differ by a field redefinition.
For {\em on-shell} matrix elements this makes no difference and as
a consequence both analyses agree for the $K\to\pi\pi$ amplitudes.
Since neither the $K\to\pi$ transition nor the $K\to$ vacuum transition
can be allowed on-shell, if the masses are such that $K\to\pi\pi$ is
possible on-shell, we first have to correctly {\em define} what we
mean by off-shell matrix elements.

Green's functions defined by quark currents, as introduced in CHPT
by Gasser and Leutwyler \cite{GL1}, 
are well defined for all values of momenta and thus
provide a proper definition of off-shell quantities. We have shown here
that for the pseudo-scalar currents as sources the two-point function
that defines properly an off-shell $K\to\pi$ transition does depend
on the weak mass term, or the coefficient $G_8'$.

Of course, we also find that for the on-shell transition $K\to\pi\pi$,
the weak mass term does not depend on $G_8'$ to order $p^2$ as shown before
in \cite{Bernard,Sonoda,Crewther}; but it does have contributions at
higher order, see the discussion below.

The discussion of \cite{KMW1} and \cite{Leurer} indicates at which
level the weak mass term can contribute to $K \to \pi \pi$ amplitudes.
The basic argument can be phrased in terms of the strong equations of motion
for the $u$ field in (\ref{lowestorder}). 
Terms that can be removed using the equations of motion
can be described by changes in the other coefficients of the Lagrangian.
The argument of Sonoda and Georgi is that the weak mass term can be written as
a total derivative assuming $m_s\ne m_d$ (see also \cite{Leurer})
and as such does not contribute to physical amplitudes. This argument fails
in the presence of external fields. It crucially requires $s+ip={\cal M}$,
otherwise other non-derivative terms involving $s+ip-{\cal M}$ remain after 
the use of the equations of motion. That is why the argument 
fails for the non-tadpole diagrams like the
two-point functions considered here.  These extra non-derivative 
terms give the $G_8'$
contributions to our $\Pi_{ij}(q^2)$ two-point functions. 

To see what happens in the case we add external vector and axial vector fields,
let us show the argument of Sonoda and Georgi in more detail
extended to include vector and axial-vector  external fields.
The basic underlying argument is that the effects of terms that vanish
using the lowest order equation of motion can be described by changes
in the other parameters of the effective Lagrangian. As we will show below
this implies that the effects of the weak mass term
can be absorbed in shifts
of the other parameters for all processes involving on-shell pseudo-scalars
and photons. As a consequence the contributions from the $G_8^\prime$ term
vanish at order $p^2$ for these type of diagrams and can be absorbed in shifts
of the other parameters at higher order, i.e. the $E_i$ and $D_i$
of \cite{KMW1} at order $p^4$. As stressed earlier, this does not mean 
that the contributions from the weak mass term are zero in this case,
only that they can be described by shifts of the other parameters.

The equation of motion from the lowest order
Lagrangian in (\ref{lowestorder}) is
\be
\label{EOM1}
2 D_\mu (U^\dagger D^\mu U)-U^\dagger\chi+\chi^\dagger U
-\frac{1}{3}\tr(-U^\dagger\chi+\chi^\dagger U) =0\,.
\ee
When the external scalar and pseudoscalar fields are zero this becomes for
$i\ne j$
\ba
\label{EOM2}
2 D_\mu (U^\dagger D^\mu U)_{ij}-2 B_0U^\dagger_{ij}m_j+2 B_0 U_{ij}m_i& =& 0
\nonumber\\
-2 D_\mu (U D^\mu U^\dagger)_{ij}-2 B_0U^\dagger_{ij}m_i+2 B_0 U_{ij}m_j&
=&0\, .
 \ea
The second equation can be derived by first multiplying
Eq. (\ref{EOM1}) on the left by $U$ and on the right by $U^\dagger$ and using
the unitarity of $U$ extensively.
The weak mass term for $\Delta S=1$ transitions is proportional to
\ba
\label{weakfinal}
(\chi^\dagger U+U^\dagger\chi)_{23}&=&
2 B_0 \left(m_d U_{23}+m_s  U^\dagger_{23}\right)
\nonumber\\
&=& 2\frac{m_s^2+m_d^2}{m_s^2-m_d^2}(D_\mu (U^\dagger D^\mu U))_{23}
+ 4\frac{m_s m_d}{m_s^2-m_d^2} (D_\mu (U D^\mu U^\dagger))_{23}\,.
\nonumber\\
\ea
For external vector and axial-vector fields zero or equal to the photon
field the last line is a total derivative and thus does not contribute
to the action. This proves the comments made above.

\begin{figure}
\begin{center}
\leavevmode\epsfxsize=12cm\epsfbox{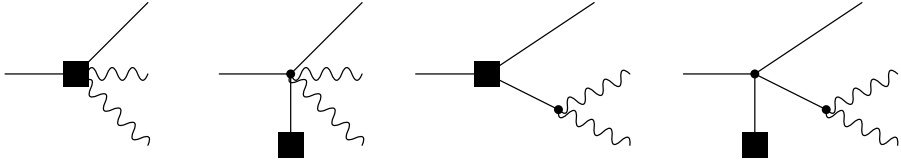}
\end{center}
\caption{The Feynman diagrams for $K_S \to \pi^0 \gamma \gamma$
at order $p^4$. Notation as in previous figures. The strong vertices
can now also be the Wess-Zumino term.\label{figure4}}
\end{figure}

Processes with on-shell pions and kaons
and photons could in principle depend on $G_8'$ already at their lowest
non-zero order.
We have verified that $G_8'$ does contribute to $K_L \to \gamma \gamma$,
and hence to $K_L \to \mu^+ \mu^-$,
at lowest non-zero order (i.e. at order $p^6$)
 similarly to the part from $G_8$\footnote{See \cite{Donoghue} for
the standard discussion and the warning about SU(3) breaking parameters}.
For $K_S \to \pi^0 \gamma \gamma$ there is already a contribution at order
$p^4$. The relevant diagrams are depicted in Fig. \ref{figure4}.
The contributions from the weak mass term are non-zero but local,
the non-local parts containing $1/(q^2_{\gamma\gamma}-m_M^2)$ cancel, 
and thus as
proved in general above can be absorbed in shifts of the $E_i$.

These results are important since 
this contribution  has never been included in the long
distance estimates of the processes above. Notice also that
the value of $G_8'$  is unknown. 
 
In addition, physical amplitudes like $K \to \pi \pi$ 
that do not depend on $G_8'$ at lowest order
can depend on it at next-to-leading order in CHPT.
In this case the $G_8'$ contribution can be 
reabsorbed in shifts of order $p^4$ and higher couplings.
Nevertheless notice that in analytical
predictions, the value of $G_8'$ should be known 
to do this shift.

\section{Conclusions}
\setcounter{equation}{0}
\label{conclusions}

We have extended the $p^2$ analysis of \cite{Bernard} with the
non-analytic contributions of \cite{Bloops} in two ways:
first, we have included all the relevant $p^4$ couplings and the full
one loop contributions to the two-point functions $K\to \pi , \eta_8$ and the 
$K\to \pi\pi$ amplitude and second, we have changed from the vague notion of an
off-shell meson field to the well-defined notion of Green's functions
of external fields in the presence of the weak non-leptonic interaction.

We have confirmed the results of \cite{Bernard} that
the three relevant couplings at order $p^2$ can be fully determined 
and discussed the necessity
of including the weak mass term in this analysis.

We concluded that the weak mass term
can contribute in some physical amplitudes 
even at their lowest non-zero order and have given several examples
where this happens. In general this can happen in the case of on-shell
Green's functions which receive contributions from
off-shell flavour changing two-point functions. As an example,
the weak mass term gives  lowest non-zero order unknown 
long-distance contributions
to processes like $K_L \to \gamma \gamma$, $K_S\to\pi^0\gamma\gamma$
and $K_L \to \mu^+ \mu^-$. 

To order $p^4$ in the chiral expansion we find six more parameters in 
the 27-plet sector
that could in principle contribute to $K\to\pi,\eta_8,\pi\pi$. Of these,
we can directly determine five from the two-point functions, leaving
one, $D_7$ as a free parameter. This parameter vanishes in the
large $N_c$ limit and is proportional to $L_4$ in the factorization model.
If the predictions of this model turn out to be satisfied by the 
other five parameters
we can take the factorization prediction for $D_7$ and obtain a 
value for $K\to\pi\pi$.

In the octet sector, there are ten more operators of which we can also
directly determine six combinations. We can use them to test the predictions of
various models like factorization, the weak deformation model \cite{EKW},
resonance models, etc.
The weaker assumptions of resonance saturation by vector, axial-vector and
scalar resonance exchange allows to determine one more combination of
counterterms from the
two-point functions and one more parameter ($E_{13}$)
can be fixed if one of the slope parameters of $K\to3\pi$, namely $K_2$ of
Ref. \cite{KMW2}, is known.
To obtain the full set of counterterms at order $p^4$ 
we need to use factorization or another more
restrictive model. Factorization and alternative models can be strongly
constrained by the value of 
the parameter combinations that can be directly extracted 
from $K\to \pi , \eta_8$ two-point functions.

As a calculational tool, we have provided the complete one-loop
 formulas for the two-point
functions of the octet symmetry case and for the on-shell 
$K\to\pi\pi$ amplitudes
with quark masses $m_s\ne m_d=m_u$ all different from zero. 
In \cite{KMW2} and \cite{EKW}
the combinations suppressed by $m_\pi^2/m_K^2$ were neglected. This might
be a good approximation for the real quark mass values 
[at the level of  10 \% though],
but will not necessarily be true on the lattice.

Since our two-point functions are much easier to determine on the lattice they
also provide a better laboratory to study unquenching effects in the
non-leptonic weak sector than the full $K\to\pi\pi$ amplitudes.

\section*{Acknowledgments}

E.P. and J.P. are grateful to the Theoretical 
Physics Department at the University of Lund for 
hospitality. The work of E.P. has been supported by Schweitzerischer
Nationalfonds. The work of J.P. has been 
partially supported by CICYT (Spain) under
Grant \# AEN96-1672. This work is also partially supported by the TMR-network,
EURODAPHNE, ERB4061PL970448. 

\appendix

\section{Order $p^4$ Loop Contributions to $\Pi_{ij}(q^2)$}
\label{applogs1}
\def\theequation{\Alph{section}.\arabic{equation}}
\setcounter{equation}{0}

Using
\ba
P(m^2) \equiv \frac{m_K^2}{16 \pi^2 F_0^2} \,
\frac{m^2}{m_K^2-m^2} \ln \left( \frac{m_K^2}{m^2} \right)
\ea
we get
\ba
 \Pi_{K^0\eta_8}(q^2) \left|_{\rm Logs}  \right. &=&
- \frac{\sqrt{Z_{K^0} Z_{\eta_8}}}
{(q^2-m_{K^0}^2)(q^2-m_{\eta_8}^2)} \, 
\frac{C}{\sqrt 6} \, \frac{F_0^4}{f_{K^0} f_{\eta_8}} \nonumber \\
& \times &   
\left[ G_8 \left\{ q^2 \left( \frac{20}{3} \mu_{\eta_8} 
+ \frac{38}{3} \mu_K  - \frac{3}{2} P(m_{\pi}^2)
+ \frac{11}{6} P(m_{\eta_8}^2) \right) \right. \right.  
\nonumber \\ &-& 
m_K^2 \left( \frac{13}{9} \mu_{\eta_8} + \frac{28}{9} \mu_K +  \mu_\pi 
+\frac{1}{2} P(m_\pi^2) +\frac{1}{18} P(m_{\eta_8}^2)
\right) \nonumber \\ &-& \left. 
m_\pi^2 \left( - \frac{1}{9} \mu_{\eta_8} + 
\frac{20}{9} \mu_K - \mu_\pi + P(m_\pi^2) + \frac{1}{9} 
P( m_{\eta_8}^2) \right) \right\} \nonumber \\ &+& 
\frac{G_{27}}{3} \left\{ - q^2 \left( 20 \mu_{\eta_8} + 78 \mu_K 
 +10 \mu_\pi + \frac{1}{2} P(m_\pi^2) + \frac{11}{2} 
P(m_{\eta_8}^2) \right) \right. \nonumber \\
&+& m_K^2 \left( \frac{13}{3} \mu_{\eta_8} - 4 \mu_K  -\frac{1}{3} \mu_\pi
- \frac{1}{6} P(m_\pi^2) +\frac{1}{6} P(m_{\eta_8}^2) \right) \nonumber \\
&+& \left. \frac{m_\pi^2}{3} \left( -\mu_{\eta_8} + \mu_\pi 
- P(m_\pi^2) + P(m_{\eta_8}^2) \right) \right\}  \nonumber \\ &-&
G_8' \left\{ q^2 \left( \frac{8}{3} \mu_8 + 6 \mu_K 
-  8 \mu_\pi - \frac{3}{2} P(m_\pi^2) + \frac{11}{6} 
P(m_{\eta_8}^2) \right) \right. \nonumber \\ 
&+& m_K^2 \left( -\frac{5}{9} \mu_{\eta_8} - \frac{20}{9}
\mu_K +\frac{5}{3} \mu_\pi - \frac{1}{2} P(m_\pi^2) -
 \frac{1}{18} P(m_{\eta_8}^2) \right) \nonumber \\ &+& 
\left. \left. m_\pi^2 \left( -\frac{1}{9} \mu_{\eta_8} 
-\frac{22}{9} \mu_K +\frac{1}{3} \mu_\pi
- P(m_\pi^2) -  \frac{1}{9} P(m_{\eta_8}^2) \right)
\right\} \right]
\ea

\ba
 \Pi_{K^0\pi^0}(q^2) \left|_{\rm Logs}  \right. &=&
- \frac{\sqrt{Z_{K^0} Z_{\pi^0}}}
{(q^2-m_{K^0}^2)(q^2-m_{\pi^0}^2)} \, 
\frac{C}{\sqrt 2}  \, \frac{F_0^4}{f_{K^0} f_{\pi^0}} \nonumber \\
& \times &    
\left[ G_8 \left\{ 
q^2 \left( \frac{8}{3} \mu_{\eta_8} 
+ \frac{26}{3} \mu_K + 8 \mu_\pi + 
\frac{1}{2} P(m_\pi^2) -  \frac{1}{6} P(m_{\eta_8}^2)
\right) \right. \right. \nonumber \\ &-& m_K^2 
\left( -\frac{1}{3} \mu_{\eta_8} + \frac{4}{9} \mu_K
 + \frac{7}{3} \mu_\pi + \frac{7}{6} P(m_\pi^2)
+ \frac{1}{18} P(m_{\eta_8}^2) \right)  
\nonumber \\ &-& \left. m_\pi^2 
\left( \frac{1}{3} \mu_{\eta_8} +\frac{8}{9} \mu_K +3  \mu_\pi 
+ \frac{1}{3} P(m_\pi^2) + \frac{1}{9} P(m_{\eta_8}^2) \right) \right\}
\nonumber \\ &+& 
\frac{G_{27}}{3} \left\{ -q^2 \left( 8 \mu_{\eta_8} + 46 \mu_K 
 +  54 \mu_\pi +  \frac{13}{2} P(m_\pi^2) - \frac{1}{2}
P(m_{\eta_8}^2) \right) \right.  
\nonumber \\ &+& m_K^2 \left( - \mu_{\eta_8} + \frac{4}{3} \mu_K  
-3 \mu_\pi - \frac{3}{2} P(m_\pi^2) + \frac{1}{6} 
P(m_{\eta_8}^2) \right) \nonumber \\
&+& \left. m_\pi^2 \left( \mu_{\eta_8} +  \frac{8}{3} \mu_K  - \mu_\pi 
+ P(m_\pi^2) + \frac{1}{3} P(m_{\eta_8}^2) \right) \right\} 
\nonumber \\ &-& G_8' \left\{ q^2 \left( \frac{2}{3} \mu_K 
+ \frac{1}{2} P(m_\pi^2) - \frac{1}{6} P(m_{\eta_8}^2) \right) 
\right. \nonumber \\ 
&+& m_K^2 \left( \frac{1}{3} \mu_{\eta_8} -\frac{4}{9}
\mu_K - \frac{7}{3} \mu_\pi - \frac{7}{6}  P(m_\pi^2)
- \frac{1}{18} P(m_{\eta_8}^2) \right)
\nonumber \\ &+&  \left. \left. m_\pi^2 \left( -\mu_{\eta_8} 
- \frac{26}{9} \mu_K + 3 \mu_\pi -  \frac{1}{3} P(m_\pi^2)
- \frac{1}{9} P(m_{\eta_8}^2) \right) \right\} \right] 
\ea

\ba
 \Pi_{K^+\pi^+}(q^2) \left|_{\rm Logs}  \right. &=&
- \frac{\sqrt{Z_{K^+} Z_{\pi^+}}}
{(q^2-m_{K^+}^2)(q^2-m_{\pi^+}^2)} \, C  \, \, \frac{F_0^4}{f_{K^+} f_{\pi^+}} 
\nonumber \\
& \times &  
\left[ G_8 \left\{ 
- q^2 \left( \frac{8}{3} \mu_{\eta_8} 
+ \frac{26}{3} \mu_K + 8 \mu_\pi 
+ \frac{1}{2} P(m_\pi^2) - \frac{1}{6} P(m_{\eta_8}^2)
\right)  \right. \right. \nonumber \\ &+& m_K^2 
\left( -\frac{1}{3} \mu_{\eta_8} + \frac{4}{9} \mu_K
 + \frac{7}{3} \mu_\pi +  \frac{7}{6} P(m_\pi^2) +
\frac{1}{18} P(m_{\eta_8}^2)  \right) \nonumber \\ &+& 
\left.  m_\pi^2 \left( \frac{1}{3} \mu_{\eta_8} + \frac{8}{9} \mu_K +3  \mu_\pi 
+ \frac{1}{3} P(m_\pi^2) + \frac{1}{9} P(m_{\eta_8}^2)
\right) \right\} \nonumber \\ &+& 
\frac{G_{27}}{3} \left\{ q^2 \left( - 7 \mu_{\eta_8} - 
 34 \mu_K - 31 \mu_\pi -  \frac{7}{2} P(m_\pi^2)- 
\frac{1}{2} P(m_{\eta_8}^2) \right) \right. \nonumber \\
&+& m_K^2 \left( \mu_{\eta_8} - \frac{4}{3} \mu_K  
-\frac{1}{3} \mu_\pi -  \frac{1}{6} P(m_\pi^2) 
 - \frac{1}{6} P(m{\eta_8}^2) \right) \nonumber \\
&+& \left. m_\pi^2 \left( - \mu_{\eta_8} + 
\frac{2}{3} \mu_K  + \mu_\pi + \frac{2}{3} 
P(m_\pi^2)  - \frac{1}{3} P(m_{\eta_8}^2) \right) \right\} 
\nonumber \\ &+&
G_8' \left\{ q^2 \left(\frac{2}{3} \mu_K 
+\frac{1}{2} P(m_\pi^2) - \frac{1}{6} P(m_{\eta_8}^2)
\right)  \right. \nonumber \\ 
&+& m_K^2 \left( \frac{1}{3} \mu_{\eta_8} - \frac{4}{9}
\mu_K - \frac{7}{3} \mu_\pi - \frac{7}{6} P(m_\pi^2) 
 - \frac{1}{18} P(m_{\eta_8}^2)  \right)
\nonumber \\ &+&  \left. \left. m_\pi^2 \left( -\mu_{\eta_8} 
- \frac{26}{9} \mu_K + 3 \mu_\pi - \frac{1}{3} P(m_\pi^2) 
- \frac{1}{9} P(m_{\eta_8}^2)  \right) \right\} \right]
\ea

\ba
 \Pi_{K^0\overline K^0}(q^2) \left|_{\rm Logs}  \right. &=&
- \frac{Z_{K^0}}{(q^2-m_{K^0}^2)^2} \, C_{\Delta S=2}
\, G_{27} \, \nonumber \\
&\times& 4 \frac{F_0^4}{f_{K^0}^2} \left[
- q^2 \left[ \frac{m_K^2}{16\pi^2 F_0^2} + 10 \mu_K
 + \frac{7}{2} \mu_\pi + \frac{9}{2} 
\mu_{\eta_8} \right]  \right. \nonumber \\ &+& \left. 2 m_K^2 \mu_K
 - \frac{1}{2} m_\pi^2 \mu_\pi 
-\frac{3}{2} m_{\eta_8}^2 \mu_{\eta_8} 
\right ] \, . 
\ea

\section{ Order $p^4$ Loop Contributions to 
$K \to \pi \pi$ Amplitudes}
\label{applogs2}
\setcounter{equation}{0}

In addition to the definitions used above we need now, 
\ba
B(m_1^2,m_2^2,p^2) &=&
\frac{1}{16 \pi^2 F_0^2} \left[ -1 + \ln \left( \frac{m_2^2}{\nu^2}
\right) + \frac{1}{2} \ln \left( \frac{m_1^2}{m_2^2}\right)
\left( 1+ \frac{m_1^2}{p^2} - \frac{m_2^2}{p^2} \right)
\right. \nonumber \\
&+& \left. {1\over 2}
\lambda^{1/2}\left( 1, \frac{m_1^2}{p^2}, \frac{m_2^2}{p^2} \right)
\ln \left[ \frac{p^2 - m_1^2 - m_2^2 
+ \lambda^{1/2}\left( p^2, m_1^2, m_2^2 \right)}
{p^2 -m_1^2 - m_2^2 -\lambda^{1/2}\left( p^2, m_1^2, m_2^2 \right)} 
\right] \right] \, , 
\nonumber 
\ea
for $p^2 > (m_1+m_2)^2$  and $p^2 \le (m_1-m_2)^2$  while
\ba
B(m_1^2,m_2^2,p^2) &=&
\frac{1}{16 \pi^2 F_0^2} \left[ -1 + \ln \left( \frac{m_2^2}{\nu^2}
\right) + \frac{1}{2} \ln \left( \frac{m_1^2}{m_2^2}\right)
\left( 1+ \frac{m_1^2}{p^2} - \frac{m_2^2}{p^2} \right)
\right. \nonumber \\
&&\hskip-2.5cm - \left. {1\over 2}
\sqrt{-\lambda\left( 1, \frac{m_1^2}{p^2}, \frac{m_2^2}{p^2} \right)}
\arctan \left[ \frac{(p^2-m_1^2-m_2^2) 
\sqrt{- \lambda(p^2,m_1^2,m_2^2)}}{(p^2-m_1^2-m_2^2)^2-2 m_1^2 m_2^2}
\right] \right]
\nonumber 
\ea
for $(m_1-m_2)^2  <  p^2 \le (m_1+m_2)^2$ and
$-\pi/2 < \arctan \, (x) < \pi/2$  
\ba
\lambda(x,y,z) &=& (x+y-z)^2 -4 xy \, . 
\ea

The non-analytic parts of $A_0^8$, $A_0^{27}$,
and $A_2$ are
\ba
\Im m \, A_0^8 \left|_{\rm Logs}  \right. &=&
- C \, G_{8} \sqrt 6 \, 
\frac{F_0^4}{f_K f_\pi^2} \,
\left[ m_\pi^2 \left(2\mu_K-\mu_\pi-\mu_{\eta_8}\right)
\right. \nonumber \\  
&+& (m_K^2-m_\pi^2) \left\{ \frac{m_K^2}{2 m_\pi^2} \left( \mu_\pi
- \mu_{\eta_8} \right) -  \frac{5}{2}  
\mu_\pi - \frac{1}{2} \mu_{\eta_8} - 3 \mu_K  \right\} \nonumber \\ 
&+& (m_K^2-m_\pi^2) \left( \frac{1}{2} m_\pi^2 
\left[ B(m_\pi^2, m_\pi^2, m_K^2)
- \frac{1}{9} B(m_{\eta_8}^2, m_{\eta_8}^2, m_K^2) \right] 
\right. \nonumber \\ &+&
\frac{1}{4} \frac{m_K^4}{m_\pi^2} \left[ B(m_K^2,m_\pi^2, m_\pi^2)
+ \frac{1}{3} B(m_K^2,m_{\eta_8}^2, m_\pi^2) \right]
\nonumber \\ &-& \left. \left. 
 m_K^2 \left[ B(m_\pi^2, m_\pi^2, m_K^2) + 
B(m_K^2, m_\pi^2, m_\pi^2) \right] \right) \right] \, ;
\nonumber \\
\ea

\ba
\Im m \, A_0^{27} \left|_{\rm Logs}  \right. &=&
- C \, G_{27} \frac{\sqrt 6}{9} \, 
\frac{F_0^4}{f_K f_\pi^2} \,
\left[ m_\pi^2 \left(2\mu_K-\mu_\pi-\mu_{\eta_8}\right)
\right. \nonumber \\  
&+& (m_K^2-m_\pi^2) \left\{ \frac{m_K^2}{2 m_\pi^2} \left( \mu_\pi
+4 \mu_{\eta_8}- 5 \mu_K  \right) -  
\frac{5}{2} \mu_\pi -\frac{11}{2} \mu_{\eta_8}- 8  \mu_K \right\} 
\nonumber \\  &+& (m_K^2-m_\pi^2) 
\left( \frac{1}{2} m_\pi^2 \left[ B(m_\pi^2, m_\pi^2, m_K^2)
+ B(m_{\eta_8}^2, m_{\eta_8}^2, m_K^2) \right] 
\right. \nonumber \\ &+& 
\frac{1}{4} \frac{m_K^4}{m_\pi^2} \left[ B(m_K^2,m_\pi^2, m_\pi^2)
- \frac{4}{3} B(m_K^2,m_{\eta_8}^2, m_\pi^2) \right] 
\nonumber \\ &-& \left. \left. 
 m_K^2 \left[ B(m_\pi^2, m_\pi^2, m_K^2) +
B(m_K^2, m_\pi^2, m_\pi^2) \right] \right) \right] \, ;
\nonumber \\
\ea

\ba
\Im m \, A_2 \left|_{\rm Logs}  \right. &=&
- C \, G_{27} \frac{10 \sqrt 3}{9} \, 
\frac{F_0^4}{f_K f_\pi^2} \,
\Bigg[ m_\pi^2 \left(2 \mu_\pi-\mu_{\eta_8}-\mu_K\right)
\nonumber \\  
&+& (m_K^2-m_\pi^2) \left\{ \frac{m_K^2}{4 m_\pi^2} \left( 5 \mu_\pi
- \mu_{\eta_8}-4\mu_K\right) 
-  10 \mu_\pi- \mu_{\eta_8} - 5 \mu_K\right\} 
\nonumber \\ &+& (m_K^2-m_\pi^2) \Bigg( -m_\pi^2 B(m_\pi^2, m_\pi^2, m_K^2)
\nonumber \\ &+& 
\frac{1}{8} \frac{m_K^4}{m_\pi^2} \left[ 5 B(m_K^2,m_\pi^2, m_\pi^2)
+ \frac{1}{3} B(m_K^2,m_{\eta_8}^2, m_\pi^2) \right]
\nonumber \\ &+&  
\frac{1}{2} m_K^2 \left[ B(m_\pi^2, m_\pi^2, m_K^2) -
2 B(m_K^2, m_\pi^2, m_\pi^2) \right] \Bigg) \Bigg] \, ;
\nonumber \\
\ea

\ba
\Re e \, A_0^8  &=& - C \, G_{8} \sqrt 6 \, 
\frac{F_0^2}{f_K f_\pi^2} \, \frac{m_K^2-m_\pi^2}{64 \pi}
\, (m_\pi^2-2m_K^2) \sqrt{1-\frac{4m_\pi^2}{m_K^2}} \, ; 
\ea

\ba
\Re e \, A_0^{27}  &=& - C \, G_{27} \frac{\sqrt 6}{9} \, 
\frac{F_0^2}{f_K f_\pi^2} \, \frac{m_K^2-m_\pi^2}{64 \pi}
\, (m_\pi^2-2m_K^2) \sqrt{1-\frac{4m_\pi^2}{m_K^2}} \, ; 
\ea

\ba
\Re e \, A_2  &=&
- C \, G_{27} \frac{10 \sqrt 3}{9} \, 
\frac{F_0^2}{f_K f_\pi^2} \, \frac{m_K^2-m_\pi^2}{64 \pi}
\, (m_K^2-2m_\pi^2) \sqrt{1-\frac{4m_\pi^2}{m_K^2}} \, ; 
\ea

\ba
\delta_i &=& -\arctan \left (\frac{\Re e \, A_i}
{\Im m \, A_i}\right ) \, \hspace*{2 cm} {\rm for } \hspace*{2 cm} 
i = 0, 2 \, . 
\ea

\section{The 27-plet Weak Lagrangian from Resonance Exchange 
Saturation}
\label{appres}
\setcounter{equation}{0}

We derive the weak effective Lagrangian at order $p^4$ in the 27-plet sector
for $\Delta S=\pm 1$ transitions by assuming resonance exchange 
saturation of the couplings. We restrict the derivation to those terms
listed in (\ref{27oper}).
The derivation of the octet sector can be found in \cite{EKW}. We refer the
reader to Refs. \cite{EKW,EGPR} for details on the method.
The weak 27-plet Lagrangian in (\ref{lagDS1}) 
can only receive contributions from scalar (octet
$S$ and singlet $S_1$) and pseudo-scalar octet $P$  resonances. 
The relevant weak couplings of the light meson fields to 
 resonances can be written as follows:
\be
{\cal L}^{27}_R = \sum_{i=1}^3 \bar{g}_S^i K_i^S +  \sum_{i=1}^2 \bar{g}_P^i 
K_i^P + \tilde{\bar{g}}_S^1 K^{S_1}_1\, +{\mbox{h.c.}}\, ,
\label{L27R}
\ee
where
\ba
&&K^S_1 = t^{ij}_{kl}~\mbox{tr}(\Delta_{ij}S)~\mbox{tr}(\Delta_{kl}\chi_+ )\, ,
~~~K^S_2 = t^{ij}_{kl}~\mbox{tr}(\Delta_{ij}\{ S,u_\mu\})~\mbox{tr}
(\Delta_{kl}u^\mu )\, , \nonumber\\
&&K^S_3 = t^{ij}_{kl}~\mbox{tr}(\Delta_{ij} S)~\mbox{tr}
(\Delta_{kl}u_\mu u^\mu )\, , \nonumber\\
&&K^P_1 = i t^{ij}_{kl}~\mbox{tr}(\Delta_{ij}P)~\mbox{tr}(\Delta_{kl}\chi_-)\,
,~~~K^P_2 = i t^{ij}_{kl}~\mbox{tr}(\Delta_{ij}[u_\mu ,P])
~\mbox{tr}(\Delta_{kl}u^\mu )\, , \nonumber\\
&& K^{S_1} = S_1~~\mbox{tr}(\Delta_{ij}u_\mu )~\mbox{tr}(\Delta_{kl} u^\mu )
\, .
\label{res27}
\ea
Inserting the lowest order solution of the equations of motion 
 for the resonance fields in (\ref{res27}), the 27-plet weak effective
Lagrangian at order $p^4$ and order $G_F$ is given by
\be
{\cal L}^{(4)}_{27} = 
\sum_{R=S,P,S_1}{1\over M_R^2}~\mbox{tr}(J^R_s J^R_w)\, ,
\label{lagres27}
\ee
where $J^R_s,\, J^R_w$ are the strong and weak currents respectively, coupled
to the resonance $R$ at lowest chiral order $p^2$. The weak currents are
defined from (\ref{L27R}) as follows  
\be
{\cal L}^{27}_R = \mbox{tr} (S J^S_w)+\mbox{tr} (P J^P_w)+S_1 J^{S_1}_w\, ,
\ee
while the strong currents are given by \cite{EGPR}
\ba
J_s^S&=&c_d~u_\mu u^\mu +c_m~\chi_+ ~~~~ J_s^{S1}=\tilde{c}_d~\mbox{tr}(u_\mu
u^\mu )+\tilde{c}_m~\mbox{tr}(\chi_+ ) \nonumber\\
J_s^P&=&id_m~\chi_- \, .
\ea  
The contributions to the low energy 27-plet weak effective Lagrangian 
(\ref{lagres27}) are summarized in Table \ref{tableres}.

\end{document}